\documentclass[prb,letterpaper,preprint,showkeys]{revtex4}

\newcommand{\citen}[1]{\onlinecite{#1}}
\usepackage{times}
\usepackage[letterpaper,margin=1.25in]{geometry}
\usepackage{version}
\newcommand{\cm}[1]{({\small \sf #1})} 
\usepackage{amsmath} 
\usepackage{amssymb}
\usepackage{url} 
\usepackage{graphicx}
\usepackage{epstopdf}
\usepackage{bibunits}
\usepackage[hypertex]{hyperref}

\newcommand{\bibendnote}[2][x]{(#2)}

\newcommand\speccite{\cite{rossberg05:_spec}}

\includeversion{onlysubmission}\excludeversion{onlypreprint}

\def\ifUnDefinedCs#1{\expandafter\ifx\csname#1\endcsname\relax}
\ifUnDefinedCs{FreezeFont} \includeversion{nowordcount} \else
\excludeversion{nowordcount}
\renewcommand\cite[1]{\ignorespaces}
\renewcommand\citen[1]{\ignorespaces}
\renewcommand\cm[1]{\ignorespaces}
\renewcommand{\maketitle}{\ignorespaces} 
\fi

\newlength{\figurewidth}

\newlength{\naturecolumn}

\renewcommand{\cm}[1]{\ignorespaces}

\date{submitted June 26, 2005, under review}

\usepackage{mydoublespace}
\def\baselinestretch{2}

\begin{document}
\title{Food Webs: Experts Consuming Families of Experts}
\author{A. G. Rossberg, H. Matsuda, T. Amemiya, K. Itoh}

\address{Yokohama National University, Graduate School of Environment
  and Information Sciences, Yokohama 240-8501, Japan}

\begin{nowordcount}
\pdfbookmark[1]{Title Page}{title}
  \vspace*{0.3\textheight}
  \maketitle
\end{nowordcount}

\begin{bibunit}[nature]

\pdfbookmark[1]{Main Text}{main}

\begin{nowordcount}
\textbf{The question what determines the structure of natural food webs has
been listed among the nine most important unanswered questions in
ecology \cite{may99:_unansered}. It arises naturally from many
problems related to ecosystem stability and resilience
\cite{mccann00:_diversity_stability_review,%
yodzis98:_local_benguel_interactions}.  The traditional \cite{yodzis81:stability_real_ecosystems%
} view is \cite{dunne05:_model_complex_stab} that population-dynamical
stability is crucial for understanding the observed
\cite{camacho02:_robus_patter_food_web_struc,
    milo02:_networ_motif,%
    neutel02:_stabil_real_food_webs,%
    williams00:_simpl,%
    cattin04:_phylog,%
    cohen90:_commun_food_webs_AND_THEREIN%
  } structures.  But phylogeny (evolutionary history) has also been
  suggested \cite{cattin04:_phylog} as the dominant mechanism.  Here
  we show that observed topological features of predatory food webs
  can be reproduced to unprecedented accuracy by a mechanism taking
  into account only phylogeny, size constraints, and the heredity of
  the trophically relevant traits of prey and predators.  The analysis
  reveals a tendency to avoid resource competition rather than
  apparent competition
  \cite{holt94:_ecolog_conseq_shared_natur_enemies}.  In food webs
  with many parasites \cite{morris01:_field_apparnt_comp} this pattern
  is reversed.}
\end{nowordcount}

\cm{Everything that looks like this is a comment and will not appear
  in the submitted version.}

Empirical food-web data is notorious for its inhomogeneity
\cite{cohen93:_improving_food_webs}.
In particular, the large number of species interacting in habitats has
forced researchers to disregard whole subsystems or to coarsen the
taxonomic resolution \cite{cohen90:_commun_food_webs_AND_THEREIN}.
The representation of trophic interactions by the simple absence or
presence of links in topological food webs is problematic, because it
turns out that by various measures \cite{berlow04:_interac_strength}
weak links are more frequent than strong links in natural food webs,
and network structures depend on a somewhat arbitrary thresholding
among the weak links
\cite{martinez99:_effec_sampl_effor_charac_food_web_struc,%
  wilhelm03:_elementary_dyn_mod}. 
Furthermore, the use of different methods for determining links
\cite{cohen93:_improving_food_webs} might affect the result.
Our analysis takes these difficulties into account by employing a
quantitative link-strength concept, an appropriate data
standardization (see Supplementary Methods), and by reflecting the
inhomogeneity of empirical methodology in our food-web model and data
analysis.


\cm{\textbf{The model:}} Specifically, the following model (``matching
model'') describing the evolution of an abstract species pool is
employed: The foraging and vulnerability traits of each species
\cite{caldarelli98:_model_multispec_commun,drossel01:_influen_predat_prey_popul_dynam,yoshida03:_dynam_web_model}
are modeled by two sequence of ones and zeros of length $n$ (the
reader might think of oppositions such as sessile/vagile,
nocturnal/diurnal, or benthic/pelagic).  The strength of trophic links
increases (nonlinearly) with the number $m$ of foraging traits of the
consumer that match the corresponding vulnerability traits of the
resource (Figure~\ref{fig:matching}).  A trophic link is considered as
present if the number of matched traits $m$ exceeds some threshold
$m\ge m_0$.  In addition, each species is associated with a size
parameter $s$ characterizing the (logarithmic) body size of a species
($0\le s <1$).  Consumers cannot forage on species with size
parameters larger than their own by more than $\lambda$.  The model
parameter $\lambda$ ($0\le \lambda <1$) controls the amount of trophic
loops \cite{polis91:_compl_des_web} in a food web.

The complex processes driving evolution are modeled by speciations and
extinctions that occur for each species randomly at rates $r_+$ and
$r_-$, respectively \cite{raup91:_phaner_kill}. New species invade the
habitat at a rate $r_1$.  Such continuous-time birth-death processes
are well understood \cite{bailey64:_stoch_proces_CHAP_8_7}.  With
$r_+<r_-$ the steady-state average of the number of species is
$r_1/(r_--r_+)$.  For new, invading species the $2n$ traits and the
size parameter $s$ are determined at random with equal probabilities.
For the descendant species of a speciation (Figure~\ref{fig:matching}),
each vulnerability trait is flipped with probability $p_\text{v}$,
each foraging trait is flipped with probability $p_\text{f}$, and a
zero-mean Gaussian random number $\delta$ ($\mathop\mathrm{var}
\delta=D$) is added to the size parameter $s$ of the predecessor
\bibendnote[s_range]{$s=0,1$ are treated as reflecting
  boundaries\cite{rossberg05:_spec}}.  Such a random, undirected model
of macroevolution becomes plausible if one assumes the trophic niche
space to be in a kind of ``occupation equilibrium'': there are no
large voids in niche space to be filled and no niche-space regions of
particularly strong predation pressure to avoid.

The model has the adjustable parameters $r_+$, $r_-$, $r_1$,
$\lambda$, $m_0$, $p_\text{v}$, $p_\text{f}$, and $D$.  For large $n$
food-web dynamics become independent of $n$, provided $m_0$ is
adjusted such as to keep the probability $C_0$ for link strengths to
exceed the threshold constant (Supplementary Discussion).  Throughout
this work $n=256$ is used.  Figure~\ref{fig:samples} and Supplementary
Figures \ref{fig:benguela}-\ref{fig:ythan96} display the connection
matrices of randomly sampled steady-state model webs in comparison
with empirical data; a Supplementary Movie illustrates the model
dynamics.

\paragraph*{}

\cm{\textbf{Model validation:}} The model was validated by comparing
snapshots of the steady state with empirical data.  Thus, only the
relative evolution rates $r_1/r_-$ and $r_+/r_-$ matter.  We set
$r_-=1$.  The size-dispersion constant $D$ has only a weak effect on
results \bibendnote[lambda]{for not too large $\lambda$, $D$, only the
  ratio $\lambda/D^{1/2}$ is relevant\speccite, as we verified
  numerically for $D=0.05/4$} and was kept fixed at $D=0.05$.  The
remaining six parameters $r_+$, $r_1$, $\lambda$, $m_0$, $p_\text{v}$,
and $p_\text{f}$ were chosen such as to fit 14 ecologically relevant,
quantitative food-web properties to empirical data (Supplementary
Figure~\ref{fig:props}), separately for each of 17 well-studied data
sets
(maximum likelihood fits, see Supplementary Methods for properties, sources of
food-web data, and fitting procedure).  Results are listed in
Table~\ref{tab:fits}.  Each fitted parameter set required $\sim 10^6$
statistically independent Monte-Carlo simulations.

To quantify the goodness-of-fit we computed $\chi^2$ statistics
(Supplementary Methods) corresponding to the remaining $14-6=8$
statistical degrees of freedom (DOF) for each data set
(Table~\ref{tab:fits}, $\chi^2_\text{M}$).  Not all empirical food-webs
are fitted equally well.  For the three food webs labeled
\textit{Scotch Broom}, \textit{British Grassland}, and \textit{Ythan
  Estuary 2} the value of $\chi^2$ exceeds the Bonferroni-corrected
$95\%$-confidence interval $\chi^2<23.0$ (15~webs).  Discrepancies
between the remaining 14 data sets and the model, on the other hand,
are revealed only when pooling all 14 sets: $\sum \chi^2=173$ for
$112\,\mathrm{DOF}$ gives $p=2\cdot 10^{-4}$.

For comparisons, the niche model \cite{williams00:_simpl} (one of the
best description known so far) was fitted to the data using the same
procedure (Supplementary Methods), and the differences $\Delta
\mathrm{AIC}=\mathrm{AIC}_\mathrm{M}-\mathrm{AIC}_\mathrm{N}$ of the
Akaike Information Criterion for fits to the matching model
($\mathrm{AIC}_\mathrm{M}$) and the niche model
($\mathrm{AIC}_\mathrm{N}$) were computed.  This statistic takes the
fact into account that the matching model is more complex and contains
four parameters more than the niche model.  Negative
$\Delta\mathrm{AIC}$ indicate that the matching model describes the
data better than the niche model and the increased model complexity is
justified.  The value is negative for 12 out of 17 models.  In the
cases where $\Delta \mathrm{AIC}>0$ this is due to unnecessary
complexity of the matching model, and not due to a better fit of the
niche model, as a comparison of the corresponding $\chi^2$ values
(Table~\ref{tab:fits}, niche model: $\chi^2_\text{N}$) shows.  Pooling
all data yields $\sum \Delta\mathrm{AIC}=-576$ in favor of the
matching model.  A comparison of the nested hierarchy model
\cite{cattin04:_phylog,rossberg05:_web,stouffer05:_quant_patterns_webs}
with our model gives similar results ($\sum
\Delta\mathrm{AIC}\approx-1480$).


\paragraph*{}

\cm{\textbf{Discussion:}} Among the fitted model parameters some
depend just as much on methodological choices at the time of recording
the food web as on the actual ecology.  In particular the linking
probability $C_0$ directly corresponds to the threshold for link
assignment, and the invasion rate $r_1$---as a parameter determining
the web size---depends on the delineation of the habitat and the
species-sampling effort.  The degree of loopiness $\lambda$ might
depend on the particular method used to determine links empirically.
Adjusting these three parameters makes the model robust to differences
in empirical methodology.

The remaining three parameters $r_+$, $p_\text{v}$, and $p_\text{f}$
allow, at least partially, an ecological interpretation.
$r_+=r_+/r_-$ represents the fraction of species that entered the
species pool by speciations from other species in the pool, in
contrast to the remaining $1-r_+$ that entered through random
``invasions''.  The low values found for $1-r_+$ (Table~\ref{tab:fits})
indicate that evolutionary processes are essential for generating the
observed structures.

The two quantities $p_\text{v}$ and $p_\text{f}$ measure the
variability of vulnerability and foraging traits among related
species.  We typically find $p_\text{v}$ much smaller than
$p_\text{f}$ (Table~\ref{tab:fits}).  In particular,
$p_\text{v}<p_\text{f}$ in 14 of 17 data sets ($p=0.006$).  This
implies that descendant species tend to acquire resources sets
different from their ancestors but mostly share their enemies.  We
interpret this as a preference for avoiding resource competition
rather than apparent competition
\cite{holt94:_ecolog_conseq_shared_natur_enemies}: A typical consumer
is an expert for its particular set of resources (resource
partitioning), and a typically resource set consists of a few
``families'' of related species---autotrophs or, again, expert
consumers.

The three exceptional data sets with $p_\text{v}/p_\text{f}>1$ are
exactly those most difficult to fit by the matching model
(Table~\ref{tab:fits}).  Interestingly, these are also the three data
sets that contain large fractions ($>30\%$) of parasites, parasitoids,
and pathogens (PPP) in the resolved species pool.  The other data sets
are dominated by predators, grazers, and primary produces (PPP
fraction $\lesssim 5\%$).  These observations are consistent with the
expectations that (i) due to their high specialization PPP are less
susceptible to resource competition than predators
\cite{morris01:_field_apparnt_comp} and (ii) the matching model does
not describe PPP well because it assumes a size ordering which is
typical only for predator-prey interactions
\cite{warren87:_pre_pry_triang,cohen93:_body_size,%
  memmott00:_predat_size_web,warren89:_spatial_freshw_web}.
But further investigations of these points are required.  For example,
contrary to expectations, $p_\text{v}/p_\text{f}$ is close to one also
for \emph{Ythan Estuary~1}.

The matching model reproduces the empirical distributions of the
numbers of consumers and resources of species well
(Figure~\ref{fig:distributions}, Supplementary
Figures~\ref{fig:benguela}-\ref{fig:ythan96}).  Under specific
conditions (see Supplementary Discussion)---including $p_\text{v} \ll
p_\text{f}$---these become the ``universal'', scaling distributions
\cite{camacho02:_robus_patter_food_web_struc,stouffer05:_quant_patterns_webs}
characteristic for the niche model (e.g.,
Figure~\ref{fig:distributions}, \textit{Caribbean Reef}).  But the
distributions for food webs deviating from these patterns are also
reproduced (e.g., Figure~\ref{fig:distributions}, \textit{Scotch
  Broom}).  An earlier variant of the matching model
\cite{rossberg05:_web} could achieve this only under unrealistic
assumptions regarding the allometric scaling of evolution rates
\cite{rossberg05:_spec}.

\paragraph*{}

Certainly there are also features of food webs that can only be
understood by taking population dynamics explicitly into account.
But, in view of the high accuracy reached with the matching model,
careful modeling of phylogeny
\cite{caldarelli98:_model_multispec_commun,%
drossel01:_influen_predat_prey_popul_dynam,yoshida03:_dynam_web_model}
should be a good starting point for further research.

\begin{nowordcount}
  \putbib[/home/axel/bib/bibview]
\end{nowordcount}

\bigskip


\bigskip

\begin{nowordcount}
  \textbf{Acknowledgements:}\newline
    \cm{Thanks (no word count?):} The authors express their gratitude
    to N.~D.~Martinez and his group for making their food-web database
    available, to T. Yamada for providing computational resources,
    N.~Rajendran for insightful comments and discussion, and to The
    21st Century COE Program ``Bio-Eco Environmental Risk Management''
    of the Ministry of Education, Culture, Sports, Science and
    Technology of Japan for financial support.

    The authors declare that they have no competing financial interests.
    
    Correspondence and requests for materials should be addressed to
    axel@rossberg.net.
\end{nowordcount}


\newpage

\begin{nowordcount}
\begin{table}[ht]
  \caption{\textbf{Goodness of fit and fitted parameters}}
  \centering
  \begin{tabular}{@{\it}l@{\rm}r@{ }lr@{ }lrrrrrrrr}
    \rm Food-web name & \multicolumn{1}{c}{$\chi^2_\text{M}$}    &  & \multicolumn{1}{c}{$\chi^2_\text{N}$}  &  & 
 \multicolumn{1}{c}{$\Delta\mathrm{AIC}$} & \multicolumn{1}{c}{$r_1$} & \multicolumn{1}{c}{$1-r_+$} & \multicolumn{1}{c}{$\lambda$} & \multicolumn{1}{c}{$m_0$} & \multicolumn{1}{c}{$C_0$} & \multicolumn{1}{c}{$p_\text{v}$} & \multicolumn{1}{c}{$p_\text{f}$} \\
Benguela Current & 14.1 & & 14.1 & & 11.6 & 2.8\phantom{3} & 0.080\phantom{1} & 0.027 & 131 & 0.38 & 0.003 & 0.32\phantom{8}\\
Bridge Brook Lake & 12.0 & & 12.4 & & 15.2 & 1.4\phantom{5} & 0.033\phantom{8} & 0.13 & 136 & 0.17 & 0.000 & 0.068\\
British Grassland & 54.7 &*& 144.6 &*& -80.1 & 1.4\phantom{0} & 0.014\phantom{3} & 0 &139  & 0.09 & 0.014 & 0.013\\
Canton Creek & 11.3 & & 12.1 & & 6.4 & 1.7\phantom{2} & 0.033\phantom{3} & 0.001 & 141 & 0.06 & 0.006 & 0.50\phantom{0}\\
Caribbean Reef & 7.5 & & 79.1 &*& -52.8 & 0.48 & 0.0082 & 0.068 & 133 & 0.29 & 0.008 & 0.39\phantom{1}\\
Chesapeake Bay & 9.5 & & 9.6 & & 10.7 & 11.5\phantom{2} & 0.25\phantom{75} & 0.001 &138  & 0.12 & 0.000 & 0.028\\
Coachella Valley & 5.9 & & 31.9 &*& -13.0 & 1.4\phantom{8} & 0.049\phantom{7} & 0.034 &124  & 0.71 & 0.002 & 0.10\phantom{2}\\
El Verde Rainforest & 14.0 & & 337.6 &*& -295.9 & 0.76 & 0.0054 & 0.12 & 139 & 0.09 & 0.015 & 0.036\\
Little Rock Lake & 11.3 & & 85.5 &*& -46.8 & 1.3\phantom{0} & 0.0092 & 0.25 & 138 & 0.12 & 0.001 & 0.043\\
Northeast US Shelf & 11.8 & & 103.6 &*& -73.9 & 0.28 & 0.0033 & 0.005 & 131 & 0.38 & 0.009 & 0.059\\
Scotch Broom & 25.8 &*& 83.3 &*& -42.5 & 1.3\phantom{7} & 0.0067 & 0.001 & 144 & 0.03 & 0.031 & 0.006\\
Skipwith Pond & 14.3 & & 39.9 &*& -10.3 & 1.4\phantom{7} & 0.045\phantom{4} & 0.033 & 130 & 0.43 & 0.011 & 0.12\phantom{7}\\
St. Marks Seegrass & 19.3 & & 37.1 &*& -3.5 & 0.55 & 0.0095 & 0.015 & 136 & 0.17 & 0.025 & 0.18\phantom{8}\\
St. Martin Island & 7.4 &  &13.7 & & -11.3 & 5.7\phantom{9} & 0.12\phantom{92} & 0 &135  & 0.21 & 0.002 & 0.32\phantom{2}\\
Stony Stream & 14.4 & & 18.5 & & -3.5 & 0.13 & 0.0033 & $10^{-5}$ & 141 & 0.06 & 0.014 & 0.35\phantom{7}\\
Ythan Estuary 1 & 20.5 & & 42.0 &*& -2.9 & 1.0\phantom{8} & 0.010\phantom{4} & 0.0029 &140  & 0.08 & 0.033 & 0.037\\
Ythan Estuary 2 & 46.4 &*& 46.6 &*& 16.1 & 2.7\phantom{1} & 0.017\phantom{0} & 0.0022 &141  & 0.06 & 0.041 & 0.038\\
  \end{tabular}

  \begin{flushleft}
    Stars (*) denote $\chi^2$ values outside the Bonferroni-corrected
    $95\%$ confidence interval.
  \end{flushleft}

  \label{tab:fits}
\end{table}
\clearpage

\noindent\textbf{Figure Legends}

\begin{figure}[hbt]
  \centering
  \includegraphics[width=0.2\figurewidth,keepaspectratio,clip]{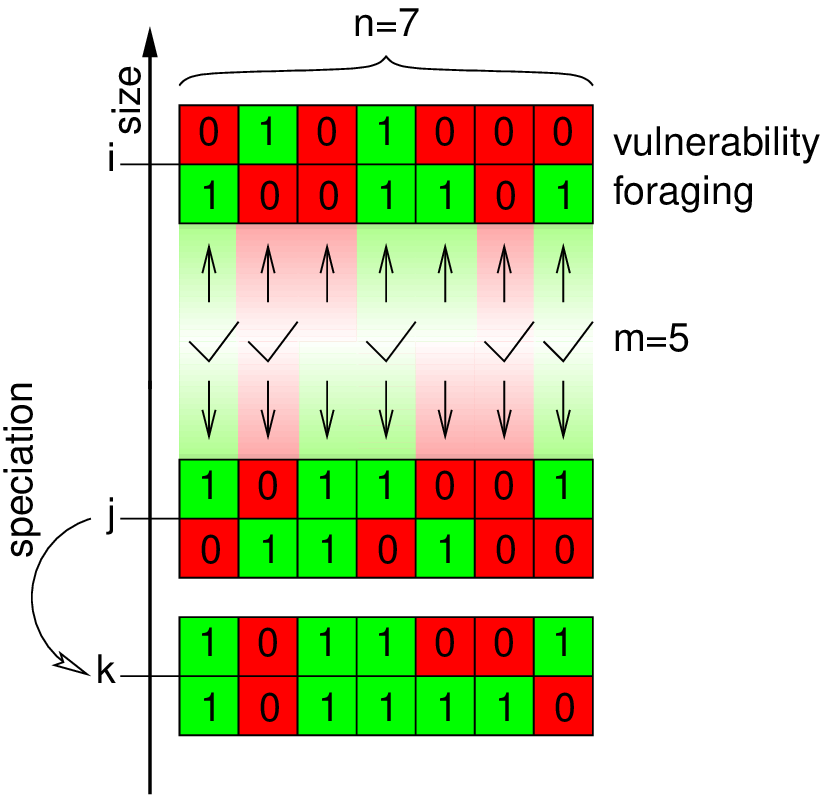}
  \caption{The main
    components of the matching model\newline Each species ($i,j,k$) is
    characterized by $n$ foraging and $n$ vulnerability traits and a
    size parameter.  Typically consumers ($i$) are larger than their
    resources ($j$).  If the number $m$ of matches between a consumer's
    foraging traits and a resource's vulnerabilities is large, trophic
    links result.  In speciations ($j\to k$) some traits mutate.
    Foraging traits typically mutate more frequently than
    vulnerability traits. See text for details.}
  \label{fig:matching}
\end{figure}

\begin{figure}[hbt]
  \centering
  \includegraphics[width=0.4\figurewidth,keepaspectratio,clip]{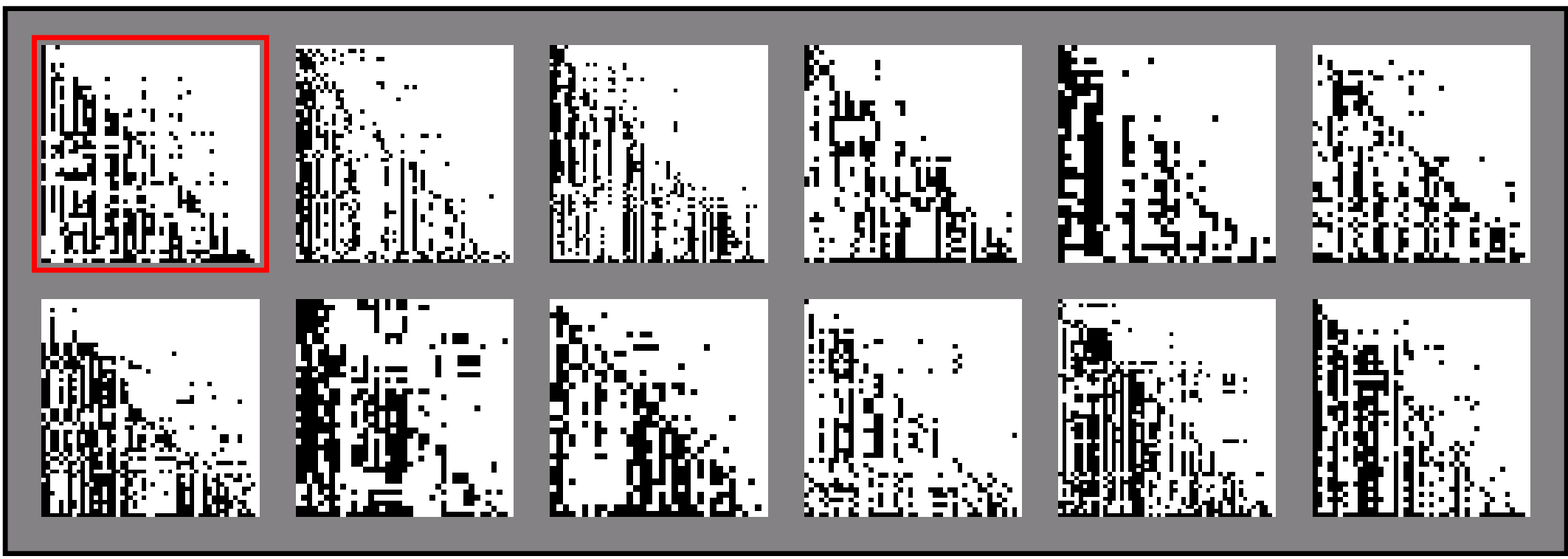}
  \caption{Comparison between model steady state and empirical
    data\newline The connection matrix of the \textit{Caribbean Reef}
    web (red box) is compared to the matrices of 11 random
    steady-state webs generated by the matching model (parameters as
    in Table~\ref{tab:fits}).  Each black pixel indicates that the
    species corresponding to its column eats the species corresponding
    to its row.  Diagonal elements correspond to cannibalism.  Pixel
    sizes vary due to varying webs sizes.  For better comparison, data
    are displayed after standardization, a random permutation of all
    species, and a subsequent re-ordering such as to minimize entries
    in the upper triangle. Characteristic are, among others, the
    vertically stretched structures \cite{cattin04:_phylog} reflecting
    the strong inheritance of consumer sets.}
  \label{fig:samples}
\end{figure}

\begin{figure}[hbt]
  \centering
  \includegraphics[width=0.3\figurewidth,keepaspectratio,clip]{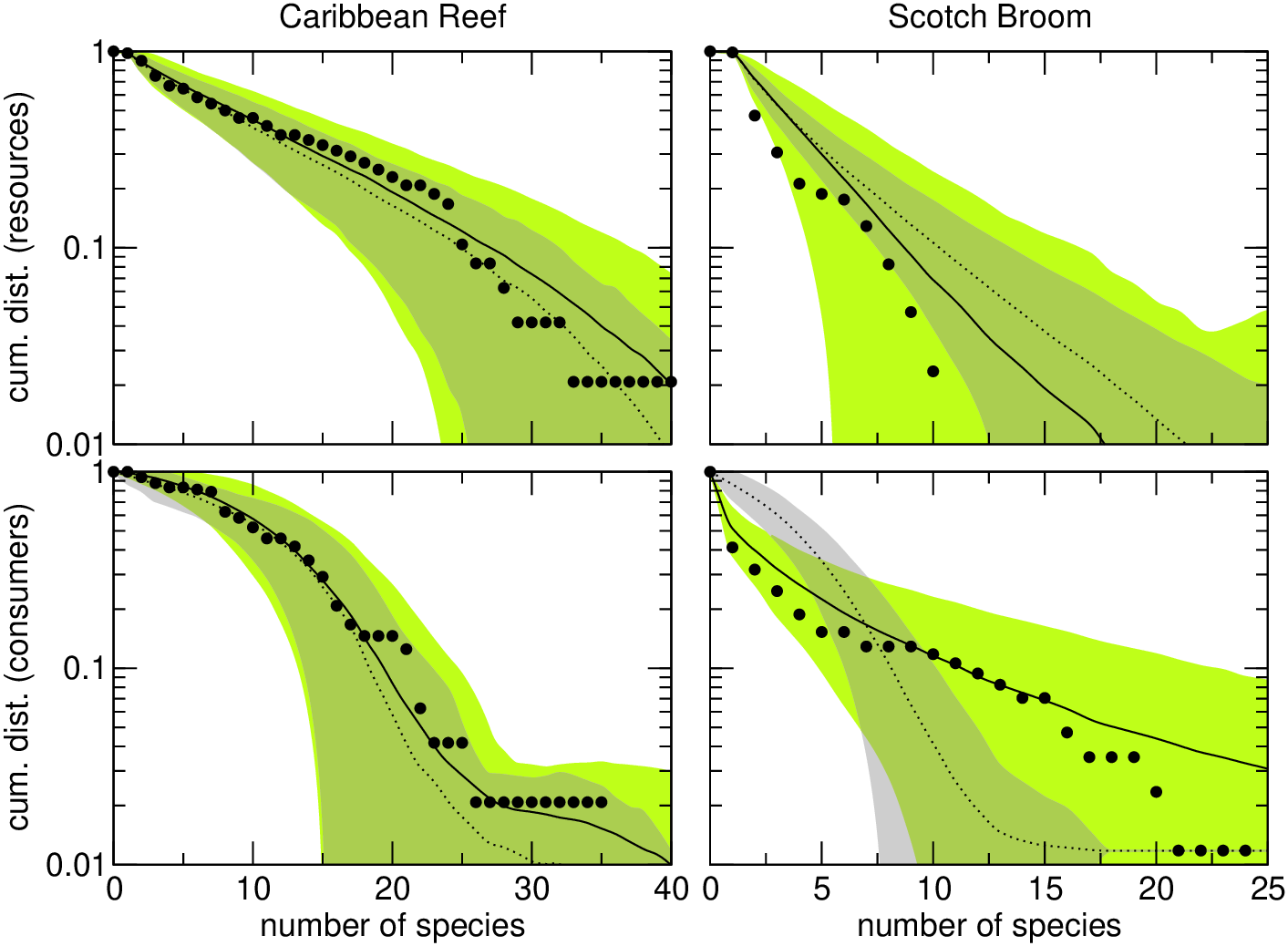}
  \caption{Food-web degree distributions\newline Cumulative
    distributions for the number of resources (upper panels) and
    consumers (lower panels) of species for the \textit{Caribbean
      Reef} and \emph{Scotch Broom} webs after data standardization.
    Points denote empirical data, solid and dotted lines model
    averages for matching and niche model, respectively.
    $2\sigma$-ranges are indicated in green (matching model) and grey
    (niche model), olive at overlaps.  Model parameters as in
    Table~\ref{tab:fits}.}
  \label{fig:distributions}
\end{figure}

\clearpage

\begin{center}
  \vspace*{\fill}
  \includegraphics[width=12cm,keepaspectratio,clip]{matching2}

  \vspace*{\fill}

  Figure 1
  \vspace*{\fill}
\end{center}
\newpage

\begin{center}
  \vspace*{\fill}
\includegraphics[width=\columnwidth,keepaspectratio,clip]{Rrow.eps}

  \vspace*{\fill}

  Figure 2
  \vspace*{\fill}
\end{center}

\newpage

\begin{center}
  \vspace*{\fill}
\includegraphics[width=\columnwidth,keepaspectratio,clip]{YR5}

  \vspace*{\fill}

  Figure 3
  \vspace*{\fill}
\end{center}

\begin{onlysubmission}
  \newpage{}
\end{onlysubmission}
\begin{onlypreprint}
  \thispagestyle{empty} \pagestyle{empty}
\end{onlypreprint}


\clearpage{}
\end{nowordcount}
\end{bibunit}

\newpage

\begin{nowordcount}

\begin{center}
  \huge  \vspace*{\fill} Supplementary Information
  \vspace*{\fill}
\end{center}


\newcommand{\mysection}[1]{\newpage%
  \renewcommand{\thesection}{#1}%
  \noindent\section*{\Large #1}%
\setcounter{subsection}{0}%
\setcounter{figure}{0}%
\setcounter{equation}{0}%
}

\begin{bibunit}[nature]
\mysection{Supplementary Methods}

\subsection{Food-Web Data}
\label{sec:data}

The food-web data base used in this work was provided by
N.~D.~Martinez and his team.  The following are references the
original sources:
\textit{Benguela Current} \cite{yodzis98:_local_benguel_interactions},
\textit{Bridge Brook Lake} \cite{havens92:_scale_webs},
\textit{British Grassland} 
\cite{martinez99:_effec_sampl_effor_charac_food_web_struc},
\textit{Canton Creek} \cite{townsend98:_distur},
\textit{Caribbean Reef} \cite{opitz96:_troph_carib},
\textit{Chesapeake Bay} \cite{baird89:_chesap_bay},
\textit{Coachella Valley} \cite{polis91:_compl_des_web},
\textit{El Verde Rainforest} \cite{waide96:_food_web_tropic_rainf},
\textit{Little Rock Lake} \cite{martinez91:_artif_attr},
\textit{Northeast US Shelf} \cite{link02:_does_web_theory},
\textit{Scotch Broom} \cite{memmott00:_predat_size_web},
\textit{Skipwith Pond} \cite{warren89:_spatial_freshw_web},
\textit{St. Marks Seegrass} \cite{christian99:_winter_seegrass},
\textit{St. Martin Island} \cite{goldwasser93:_const_carib_web},
\textit{Stony Stream} \cite{townsend98:_distur},
\textit{Ythan Estuary~1} \cite{hall91:_food_rich_web},
\textit{Ythan Estuary~2} \cite{huxham96:_do_parasites}.

\subsection{Statistical Analysis}
\label{sec:properties}

\subsubsection{Data standardization}
\label{sec:standardization}

\noindent Both empirical and model data were evaluated/compared after applying a
data standardization procedure to the raw data.  The procedure
consists of three steps:
\begin{enumerate}
\item Deleting disconnected species and small, disconnected sub-webs.
  Graph theory predicts that there will be only a single large connected
  component.  We keep only this large component.
\item Lumping of all species at the lowest trophic level into a single
  ``trophic species''.  We do this, because in some data sets the
  lowest trophic level is already strongly lumped.  For example, the
  \textit{Chesapeake Bay} web contains a species ``phytoplankton'',
  and Coachella Valley ``plants/plant products''.  On the other hand,
  food webs such as \textit{Little Rock Lake} resolves the
  phytoplankton at the genus level.  Lumping the lowest level improves
  data intercomparability.
\item The usual lumping of trophically equivalent species into single
  ``trophic species'' \cite{cohen90:_commun_food_webs_AND_THEREIN}.
\end{enumerate}
For some data sets with a simple structure, this procedure leads to a
considerable reduction of the web size (e.g., \textit{Bridge Brook
  Lake} shrinking from 74 species to 15).  But generally this is not
the case.

\subsubsection{Food-Web Properties}

Besides the number of species $S$ and the number of links $L$
expressed in terms of the directed connectance
\cite{martinez91:_artif_attr} $C=L/S^2$ , the following 12 food-webs
properties were used to characterize and compare empirical and model
webs: the clustering coefficient
\cite{camacho02:_robus_patter_food_web_struc,
  dorogovtsev02:_evolut_networ} (\textit{Clust} in Supplementary
Figure~\ref{fig:props}); the fractions of cannibalistic species
\cite{williams00:_simpl} (\textit{Cannib}) and species without
consumers \cite{cohen90:_commun_food_webs_AND_THEREIN} (\textit{T},
top predators); the relative standard deviation in the number of
resource species \cite{schoener89:_food_webs} (\textit{GenSD},
generality s.d.)  and consumers \cite{schoener89:_food_webs}
(\textit{VulSD}, vulnerability s.d.); the web average of the maximum
of a species' Jaccard similarity \cite{jaccard08:_nouvel_florale} with
any other species \cite{williams00:_simpl} (\textit{MxSim}); the
fraction of triples of species with two or more resources, which have
sets of resources that cannot be ordered to be all contiguous on a
line \cite{cattin04:_phylog} (\textit{Ddiet}); the average
\cite{cohen90:_commun_food_webs_AND_THEREIN} (\textit{aChnLg}),
standard deviation \cite{martinez91:_artif_attr} (\textit{aChnSD}),
and average per-species standard deviation
\cite{goldwasser93:_const_carib_web} (\textit{aOmniv}, omnivory) of
the length of food chains, as well as the $\log_{10}$ of their total
number \cite{martinez91:_artif_attr} (\textit{aChnNo}), with the
prefix \textit{a} indicating that these quantities were computed using
the fast, ``deterministic'' Berger-Shor approximation
\cite{berger90:_approx_subgraph} of the maximum acyclic subgraph (MAS)
of the food web. The number of non-cannibal trophic links not included
in the MAS was measured as \textit{aLoop}. When the output MAS of the
Berger-Shor algorithm was not uniquely defined, the average over all
possible outputs was used.

All food-web properties were calculated after data standardization as
described above.

\subsubsection{Goodness-of-fit statistics}

Mean and covariance matrix of the food-web properties described above,
including $C$ but not $S$, were computed for the model steady state
and projected \cite{rossberg05:_web} to $S$ fixed at the empirical
value.  The corresponding log-likelihood and the $\chi^2$ of the empirical
values were computed thereof assuming Gaussian distributions.  See
Ref.~\citen{rossberg05:_web} for details.

\subsubsection{Parameter Fitting}

The fitting parameters listed in Table~\ref{tab:fits} (except $r_1$)
were chosen such as to maximize the log-likelihood computed as
described above (maximum likelihood estimates).  Given the other
parameters, $r_1$ was always adjusted such as to make the model
expectation value of $S$, determined from Monte Carlo simulations,
match the empirical value.  The Akaike Information Criterion follows
directly from the log-likelihood of the best-fitting parameter set.

In order to compute a comparable Akaike Information Criterion for the
niche model, some modification of the original prescription for this
model \cite{williams00:_simpl} were required:
\begin{itemize}
\item We applied the data standardization
  (\ref{sec:standardization}) to \emph{both} model and empirical
  data,
\item determined the niche-model parameter \cite{williams00:_simpl}
  $\beta$, which controls the connectance, by a maximum-likelihood
  estimate as above, and
\item determined the number of species of model webs before data
  standardization such as to match the expected number of species
  after data standardization with the empirical data, just as
  described above for the parameter $r_1$ of the matching model.
\end{itemize}

\putbib[/home/axel/bib/bibview]

\end{bibunit}

\begin{bibunit}[nature]

\mysection{Supplementary Discussion}

\subsection{Derivation of Link Dynamics for Large $n$}
\label{sec:universality}

Here we explain why the network dynamics of the matching model becomes
independent of $n$ for large $n$, if $m_0$ is properly adjusted as $n$
increases.  First, consider a single trophic link from a (potential)
consumer to a (potential) resource.  Denote the foraging traits of the
former by $f_i$, the vulnerability traits of the latter by $v_i$,
where $i=1,\ldots,n$ and $f_i,v_i\in\{0,1\}$.

\subsubsection{Linking Probability}
\label{sec:C0}

Consider the steady-state distribution of the link strength $m$
defined by
\begin{align}
  \label{defm}
  m=\sum_{i=1}^n
  \left\{
      \begin{matrix}
        1&\text{if $f_i = v_i$}\\ 
        0&\text{if $f_i\neq v_i$}
      \end{matrix}
    \right\}.
\end{align}
Since the $f_i$ and $v_i$ are equally, independently distributed, $m$
follows a binomial distribution with mean $n/2$ and standard deviation
$\sigma=n^{1/2}/2$.  The probability for a link to exceed the
threshold $m_0$ is 
\begin{align}
  \label{m0}
  C_0:=P(m\ge m0)=2^{-n} \sum_{m=m_0}^n \binom{n}{m}.
\end{align}
The distribution of $x:=(m-n/2)/\sigma$ converges to a standard normal
distribution of large $n$.  The linking probability $C_0$ converges
to a fixed value $(2\pi)^{-1/2}\int_{x_0}^\infty \exp(-t^2/2)\,dt$ if $m_0$ is adjusted
such that $(m_0-n/2)/\sigma$ converges to a fixed value $x_0$.

\subsubsection{Mutation as an Integrated Ornstein-Uhlenbeck Process}
\label{sec:betas}

In the following we argue that the dynamics of $x$ between speciations
can be characterized as an integrated Ornstein-Uhlenbeck process if
$n$ is large.  First, consider only a single link, as above.  When the
resource speciates, its vulnerability traits are inherited by the
descendant species, but with probability $p_\text{v}$ they flip from
$v_i$ to $1-v_i$.  If $p_\text{v}< 1/2$ this single step can be
divided into a series of $K$ small steps, where a property $v_i$ is
flipped in each step with a small probability $q$ and otherwise left
unchanged.  Taking the possibility that properties are flipped
repeatedly into account, one finds that the $K$ small steps are
equivalent to the speciation step if
\begin{align}
  \label{pq}
  p_\text{v}=\frac{1}{2}
  \left[
    1-(1-2\,q)^K
  \right]
\end{align}
or 
\begin{align}
  \label{q}
  q=-\frac{\log(1-2\,p_\text{v})}{2\,K}+\mathcal{O}\!\left(K^{-2}\right).
\end{align}
For sufficiently large $K$ one has $q\,n \ll 1$.  Then at most one
trait is flipped in each step, and the change in $x=(m-n/2)/\sigma$ is
of order $\sigma^{-1}\sim n^{-1/2}$.  As $n$ increases, it becomes
arbitrarily small.

Denote the value of $m$ after the $k$-th step by $m_k$.  At each step,
if $m_k$ is known, the probability distribution of $m_{k+1}$ depends
only on $n$ and $m_{k}$.  If $q\,n \ll 1$, for example, one has
$m_{k+1}=m_{k}-1$ with probability $m_k\,q$, 
$m_{k+1}=m_k+1$ with probability $(n-m)\,q$, and otherwise
$m_{k+1}=m_k$.  Thus the dynamics of $m$---and of $x$---from step to
step are Markov processes.  

These three properties of the step-by-step dynamics of $x$ in the limit
of large $n$ and $K$
\begin{enumerate}
\item normal distribution in the steady state
\item Markov property
\item arbitrarily small changes from step to step
\end{enumerate}
identify the dynamics as an Ornstein-Uhlenbeck process
\cite{gardiner90:_handb_stoch_method}
\begin{align}
  \label{OU}
  dx(\tau)=-\mu\, x(\tau)\,dt+\eta\, dW(\tau),
\end{align}
where $W(\tau)$ is a Wiener process
\cite{gardiner90:_handb_stoch_method} and $\tau=k/K$.  In particular,
one finds
\begin{align}
  \label{rs}
  \mu=-\log(1-2p_\text{v}),\qquad  \eta=\sqrt{2\,\mu}.
\end{align}
The value of $x$ for a link from a speciating resource to its consumer
is given by the integral of Eq.~(\ref{OU}) over a $\tau$-interval of
unit-length, starting with the value of $x$ for the ancestor.  This
implies that of the correlation of $x$ between direct relatives is
$(1-2p_\text{v})$ and between relatives of $l$-th degree
$(1-2p_\text{v})^l$.  The corresponding results for a speciating
consumer are obtained by replacing $p_\text{v}$ in Eq.~(\ref{rs}) by
$p_\text{f}$. 

For the inheritance of several links to unrelated (hence uncorrelated)
consumers, Eq.~(\ref{OU}) holds for each link, and the Wiener
processes are uncorrelated.  For links to unrelated resources
correspondingly.  For links to related species the Wiener processes
are correlated.  From invariance considerations regarding the temporal
ordering of evolutionary events in local networks one finds that for
relatives of $l$-th degree this correlation is $(1-2p_\text{f})^l$ for
species-as-consumers and $(1-2p_\text{v})^l$ for species-as-resources.
The correlations between links to related species from a newly
invading species also follow this pattern.  This provides a full
characterization of the link dynamics for large $n$ independent of
$n$.

\subsection{Relation to Previous Analytic Results}
\label{sec:markov}

In order to make the analytic characterizations of the degree
distributions and other food-web properties obtained for an earlier
model variant \cite{rossberg05:_spec} accessible for the matching
model, we derive an approximate description of the link dynamics that
refers directly to the inheritance of connectivity between species,
i.e., of the information if a link is present or not, rather than the
inheritance of traits determining links.

Mathematically, this corresponds to a Markov approximation for the
dynamics of the connectivity in the following form: If resource B
speciates to C, its connectivity information to a consumer A is lost
with a probability $\beta_\mathtt{v}$ (independent of the previous
history) and otherwise copied from B to C.  When the information is
lost, a link from C to A is established at random with probability
$C_0$.

The breaking probability $\beta_\mathtt{v}$ can be obtained by
equating the probabilities A eats C given that A eats C's ancestor B
for the exact description (in terms of $p_\mathtt{v}$ and $m_0$) and
the Markov approximation.  This gives
\begin{align}
  \label{betav}
  \beta_\text{v}=\frac{1}{2^n\,(1-C_0)\,C_0}\sum_{m_1=m_0}^n
  \sum_{k=0}^{n-m_1} \sum_{m_2=k}^{m_0-1}
  \frac{n!\,\,p_\text{v}^{2k+m_1-m_2}\,(1-p_\text{v})^{n-2k-m_1+m_2}}{k!\,(k+m_1-m_2)!\,(m_2-k)!\,(n-m_1-k)!}
\end{align}
with $C_0$ defined by Eq.~(\ref{m0}).  The corresponding expression
for $\beta_\text{f}$ is obtained by replacing $p_\text{v}$ in
Eq.~(\ref{betav}) by $p_\text{f}$.  Results of
Ref.~\citen{rossberg05:_spec} can be applied to the matching model with
the replacement of the parameter $\beta$ in
Ref.~\citen{rossberg05:_spec} by $\beta_\text{v}$.  

Most analytic results of Ref.~\citen{rossberg05:_spec} rely on the
unrealistic assumption of consumers evolving much slower than their
resources.  This assumption is used to argue for 
\begin{enumerate}
\item fully developed correlations of connectivity from one consumer
  to related resources and
\item absence of correlations for connectivity from one resource to
  related consumers.
\end{enumerate}
Effects 1 and 2 are then used to simplify calculations.  In the
matching model 1 and 2 can be obtained without assuming large
differences in speciation rates: Effect 1 is obtained because
statistical correlations in connectivity to related resources in the
matching model depend only on the correlations between the traits of
the resources, and not on the evolutionary history of the consumer
(see also \ref{sec:universality}).  The correlations are large if
$p_\text{v}$ is small and, as a result, $\beta_\text{v}$ is small.
Effect 2 is obtained when $p_\text{f}$ is close to $0.5$ (foraging
traits are randomized in speciations), which implies that
$\beta_\text{f}$ is close to $1$.

Results of Ref.~\citen{rossberg05:_spec} that contribute to a better
understanding of the matching model include the derivation of the
conditions under which the degree distributions become those of the
niche model, and the explanation why model webs, just as empirical
data \cite{cohen90:_commun_food_webs_AND_THEREIN}, exhibit a
larger-than-random degree of ``intervality''.  The average number of
resource ``families'' (or ``clades'') of a consumer in the matching
model can also be estimated, and turns out to be small: The largest
value (3.7) is obtained for the top predator of \emph{Ythan Estuary
  2}.  For most other webs this number is below two.

\putbib[/home/axel/bib/bibview]

\end{bibunit}

\mysection{Supplementary Movie Legend}
\label{sec:movie}

(The movie can be found at \url{http://ag.rossberg.net/matching.mpg}
or \url{http://www.envcomplex.ynu.ac.jp/matching.mpg}.)

This 1 minute movie (MPEG, 7MB) illustrates the dynamics of the
matching model.  The movie shows the evolution of the connection
matrix of food webs in the model steady state at parameters
corresponding to \textit{Little Rock Lake} (Table~\ref{tab:fits}).
Each black pixel indicates that the species corresponding to its
column eats the species corresponding to its row.  Diagonal elements
correspond to cannibalism. To ensure temporal continuity, the raw
data---prior to data standardization---are show.  Thus, these matrices
are not directly comparable to the matrices displayed in
Figure~\ref{fig:samples} and Supplementary
Figures~\ref{fig:benguela}-\ref{fig:ythan96}.  Species are sorted by
decreasing size parameter $s$ from top to bottom and left to right.
The movie shows one evolutionary event (speciation, extinction,
invasion) per frame at 25 frames per second.


\begin{bibunit}[nature]

\mysection{Supplementary Figure Legends}
\label{sec:props2}

\subsection{Results for Food-Web Properties}
\label{sec:figprops}

In Supplementary Figure~\ref{fig:props} the best fitting results for
the matching model (red starts) and for the niche model (blue boxes)
are compared to the empirical data (horizontal lines).  Vertical lines
correspond to $\pm$ one model standard deviation.  Because the
properties are computed conditional to fixed $S$, the value of $S$
always fits exactly.  Note that the graph does not contain the full
information about the covariance matrices that entered the $\chi^2$
and likelihood calculations, and therefore indicates the goodness of
fit only semiquantitatively.

\subsection{More connection matrixes and degree distributions}
\label{sec:more}

Supplementary Figures~\ref{fig:benguela} to \ref{fig:ythan96} present
the results corresponding to Figures~\ref{fig:samples} and
\ref{fig:distributions} (main text) for all food webs and the two
models considered.  In each figure, the first panel shows the
connection matrix of the empirical food web in a red box compared to
the first 11 random samples obtained from a simulation of the matching
model.  As in Figure~\ref{fig:samples}, each black pixel indicates
that the species corresponding to its column eats the species
corresponding to its row.  Diagonal elements correspond to
cannibalism.  Pixel sizes vary due to varying webs sizes.  Data are
displayed after standardization, a random permutation of all species,
and a subsequent re-ordering such as to minimize entries in the upper
triangle.  The second panel in each figure displays the corresponding
data for the niche model.

The two bottom panels compare model and empirical degree distributions
(model parameters as in Table~\ref{tab:fits}).  As in
Figure~\ref{fig:distributions}, points denote empirical data, and solid
and dotted lines model averages for matching and niche model,
respectively; $2\sigma$-ranges are indicated in green (matching model)
and grey (niche model), olive at overlaps.  All model distributions
were calculated conditional to $S$ fixed at the empirical value.

Since, for the purpose of data standardization, the lowest trophic
level is lumped to a single trophic species, there is always exactly
one ``species'' that does not consume others.  As a result, the second
point in the cumulative distribution of the number of resources is
always fixed at $(S-1)/S$.  Because the consumers of this lumped
species are the consumers of all species that were lumped into it, the
number of consumers of the lumped species is comparatively large,
which leads to a leveling-off at the tails of consumer distributions
as compared to the distributions for the raw data shown in
Ref.~\citen{stouffer05:_quant_patterns_webs}.

The informations provided by the connection matrices and the degree
distributions are complementary.  While the degree distributions give
integral information regarding the whole web, the connection matrices
give an impression of the correlations present between individual
species as well as the fluctuations in the steady state.

\putbib[/home/axel/bib/bibview]

\mysection{Supplementary Figures}
\label{sfigs}

\vspace{\fill}

\begin{figure}[!h]
  \centering
  \caption{rotated for better resolution, see \ref{sec:figprops} for explanations}
  
  \bigskip

  \includegraphics[angle=-90,width=0.8\textwidth,keepaspectratio,clip]{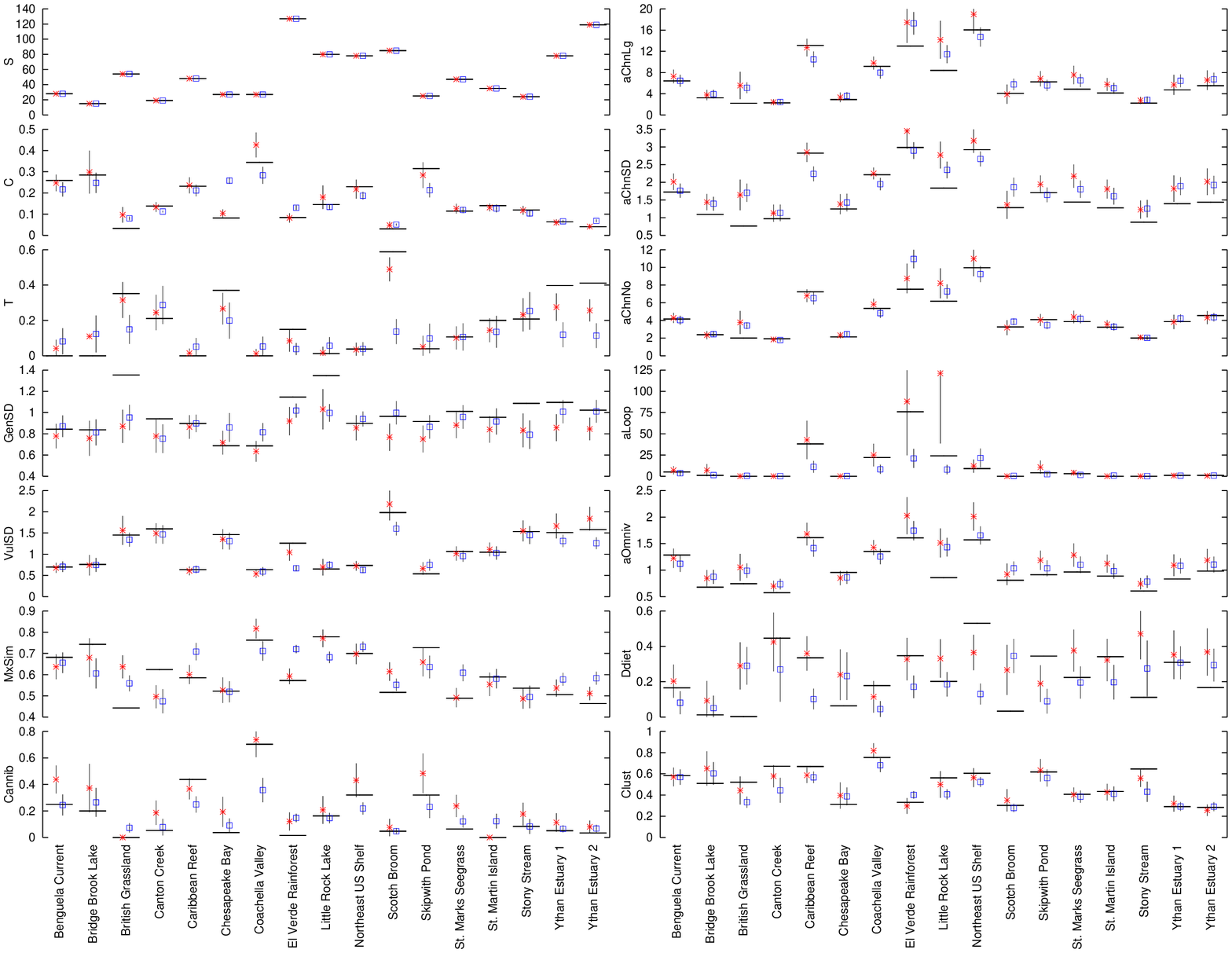}
  \label{fig:props}
\end{figure}

\vspace{\fill}

\newpage 

\begin{figure}[h]
  \caption{Benguela Current (see \ref{sec:more} for explanations)}
  \textsf{Matching model}:\\
  \includegraphics[width=2\figurewidth,keepaspectratio,clip]{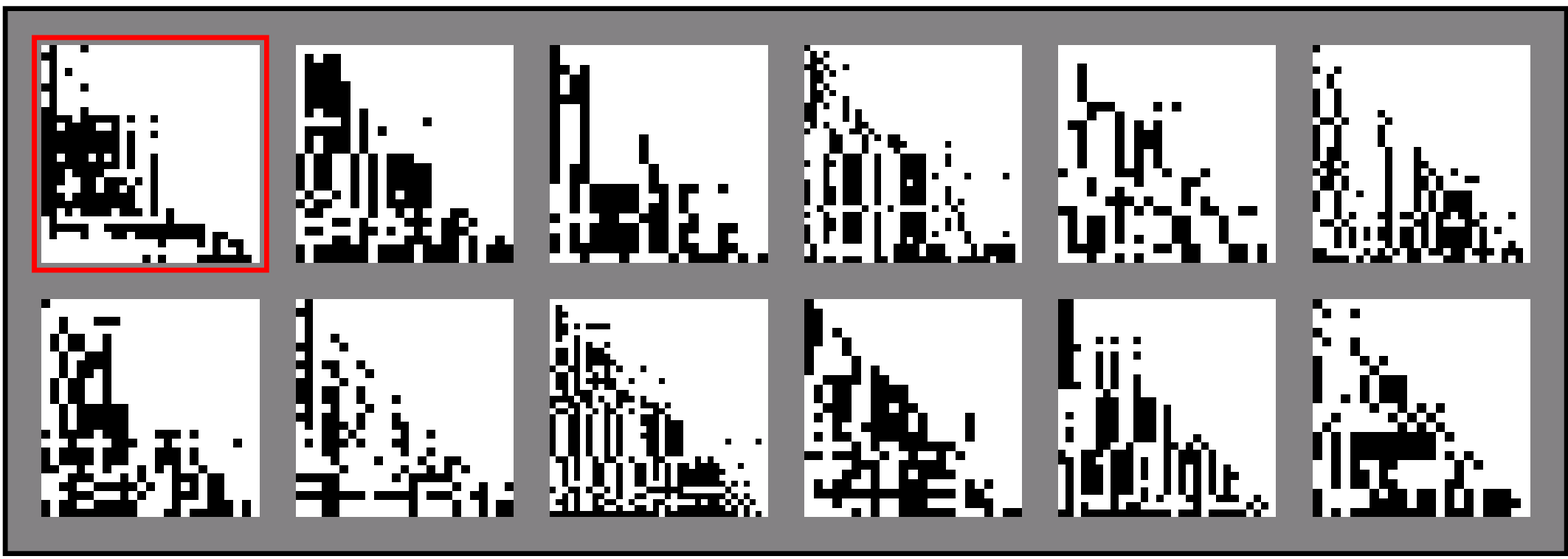}\\
  \textsf{Niche model}:\\
  \includegraphics[width=2\figurewidth,keepaspectratio,clip]{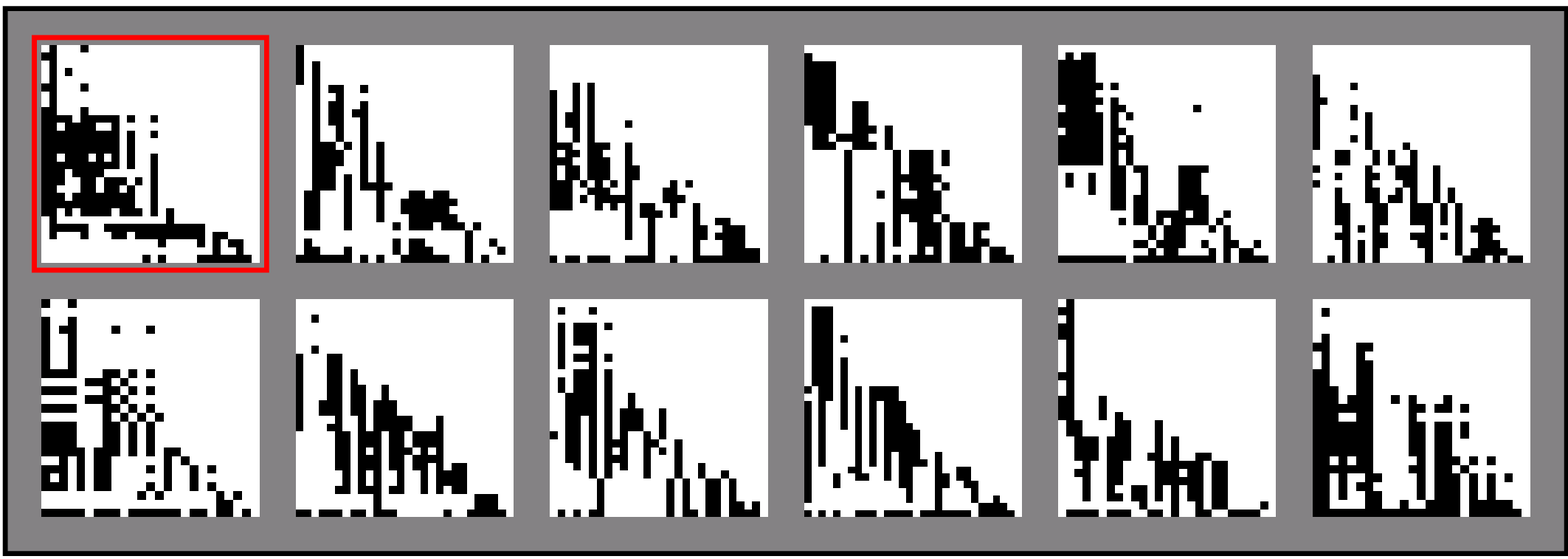}\\

\smallskip

  \includegraphics[width=2\figurewidth,keepaspectratio,clip]{Benguela-RY.eps}
  \label{fig:benguela}
\end{figure}
\begin{figure}[h]
  \caption{Bridge Brook Lake (see \ref{sec:more} for explanations)}
  \textsf{Matching model}:\\
  \includegraphics[width=2\figurewidth,keepaspectratio,clip]{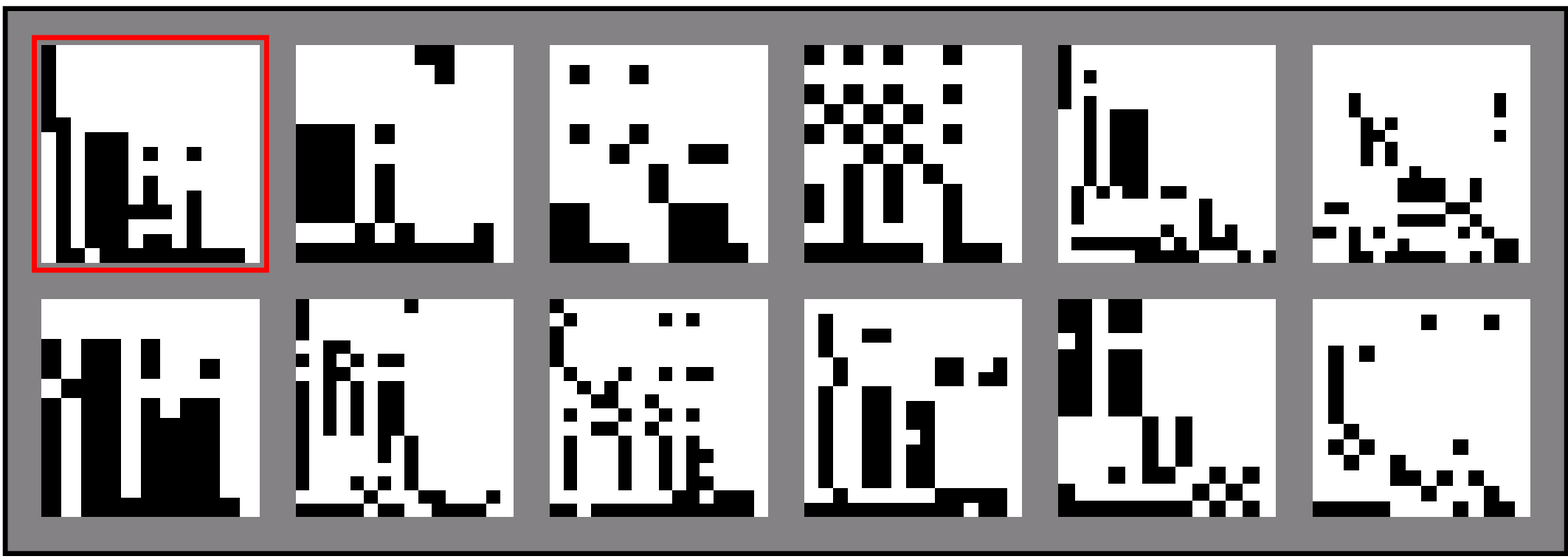}\\
  \textsf{Niche model}:\\
  \includegraphics[width=2\figurewidth,keepaspectratio,clip]{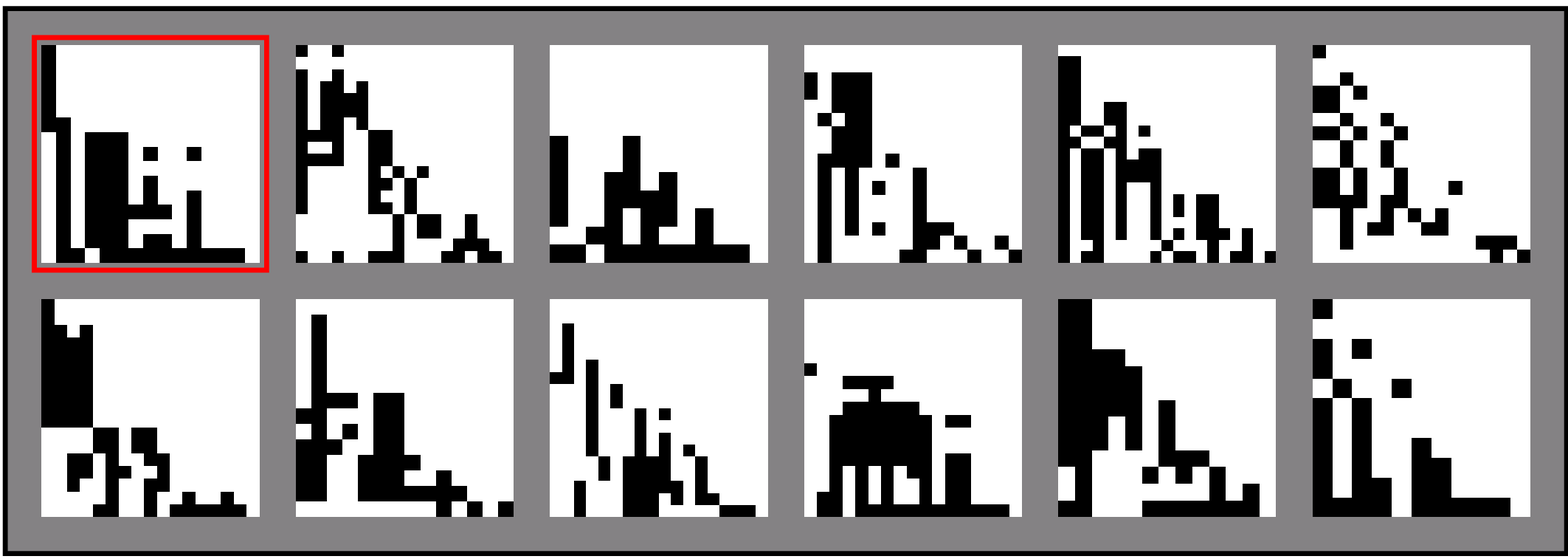}\\

\smallskip

  \includegraphics[width=2\figurewidth,keepaspectratio,clip]{Bridge-Brook-RY.eps}
  \label{fig:bridge}
\end{figure}
\begin{figure}[h]
  \caption{British Grassland (see \ref{sec:more} for explanations)}
  \textsf{Matching model}:\\
  \includegraphics[width=2\figurewidth,keepaspectratio,clip]{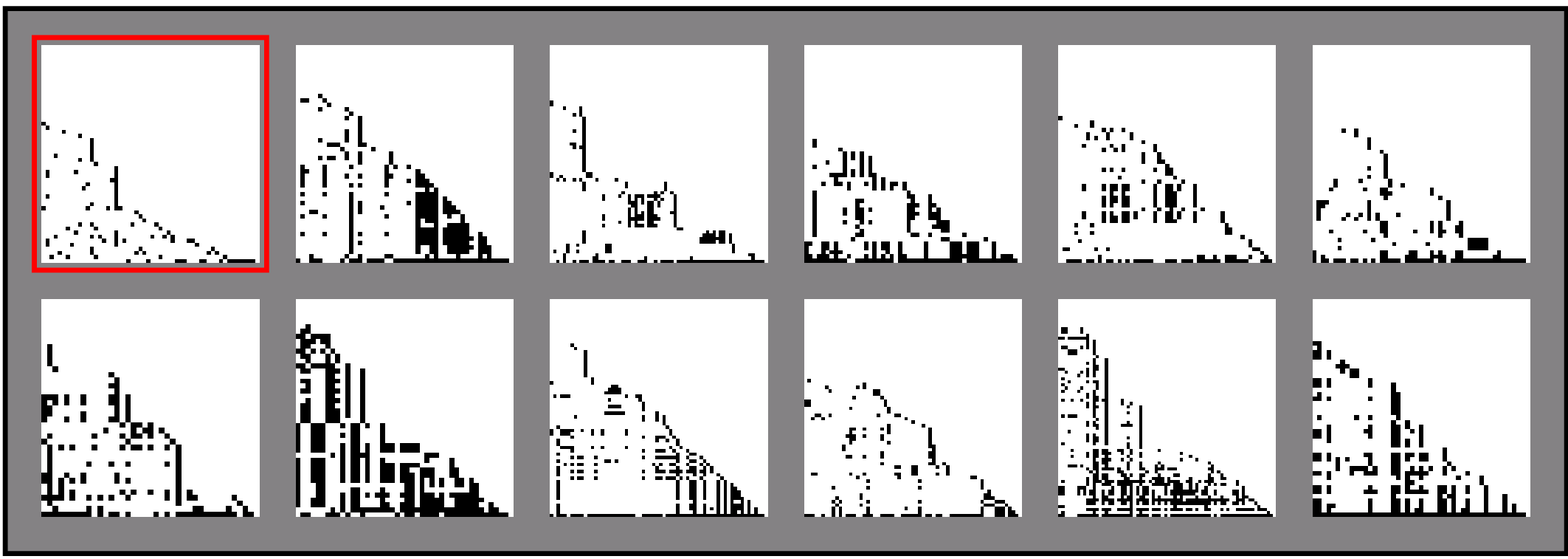}\\
  \textsf{Niche model}:\\
  \includegraphics[width=2\figurewidth,keepaspectratio,clip]{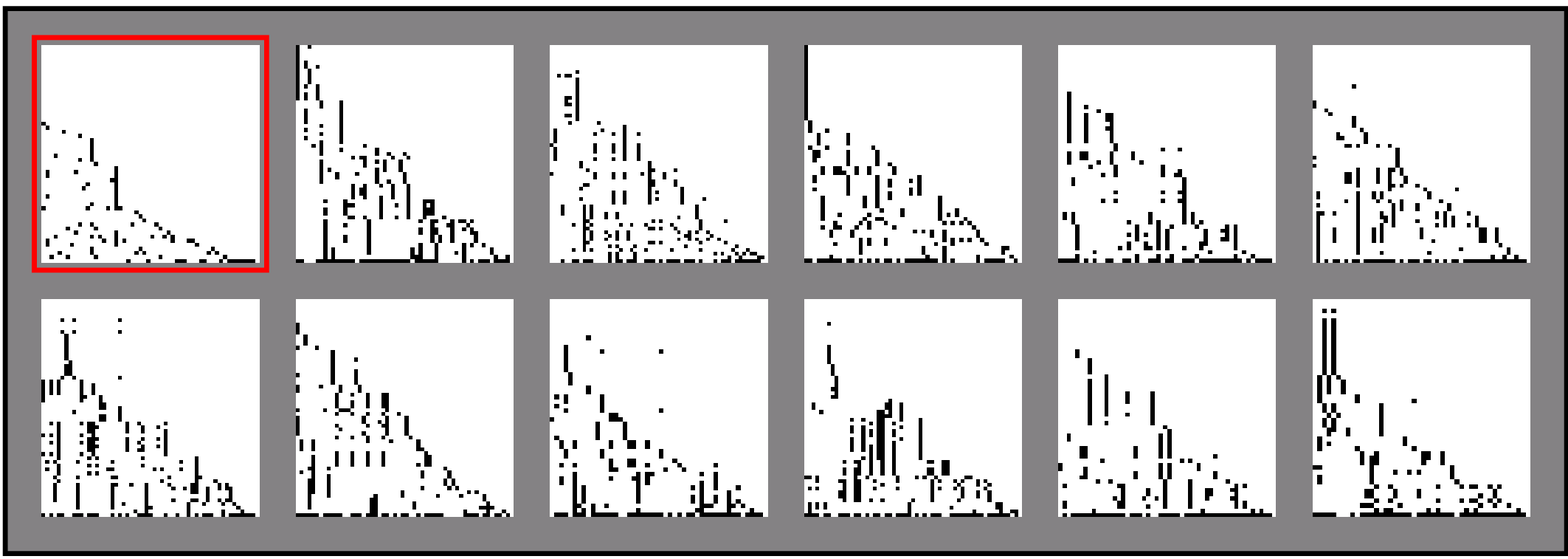}\\

\smallskip

  \includegraphics[width=2\figurewidth,keepaspectratio,clip]{Grass-RY.eps}
  \label{fig:grass}
\end{figure}
\begin{figure}[h]
  \caption{Canton Creek (see \ref{sec:more} for explanations)}
  \textsf{Matching model}:\\
  \includegraphics[width=2\figurewidth,keepaspectratio,clip]{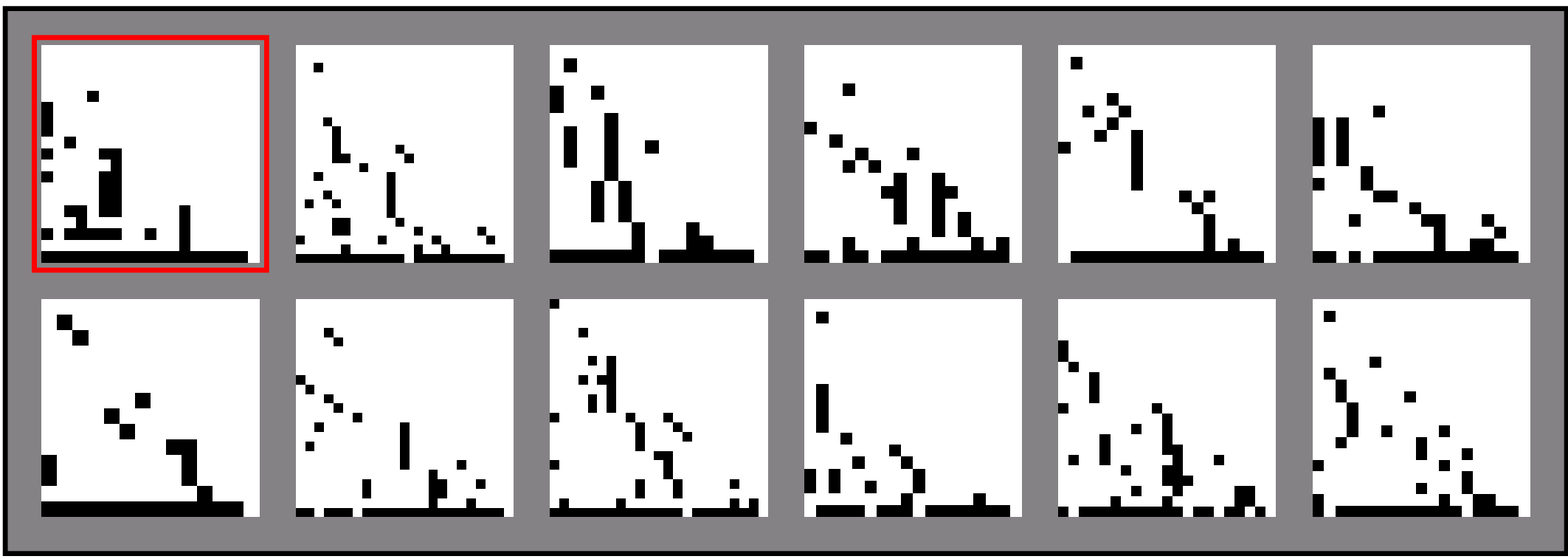}\\
  \textsf{Niche model}:\\
  \includegraphics[width=2\figurewidth,keepaspectratio,clip]{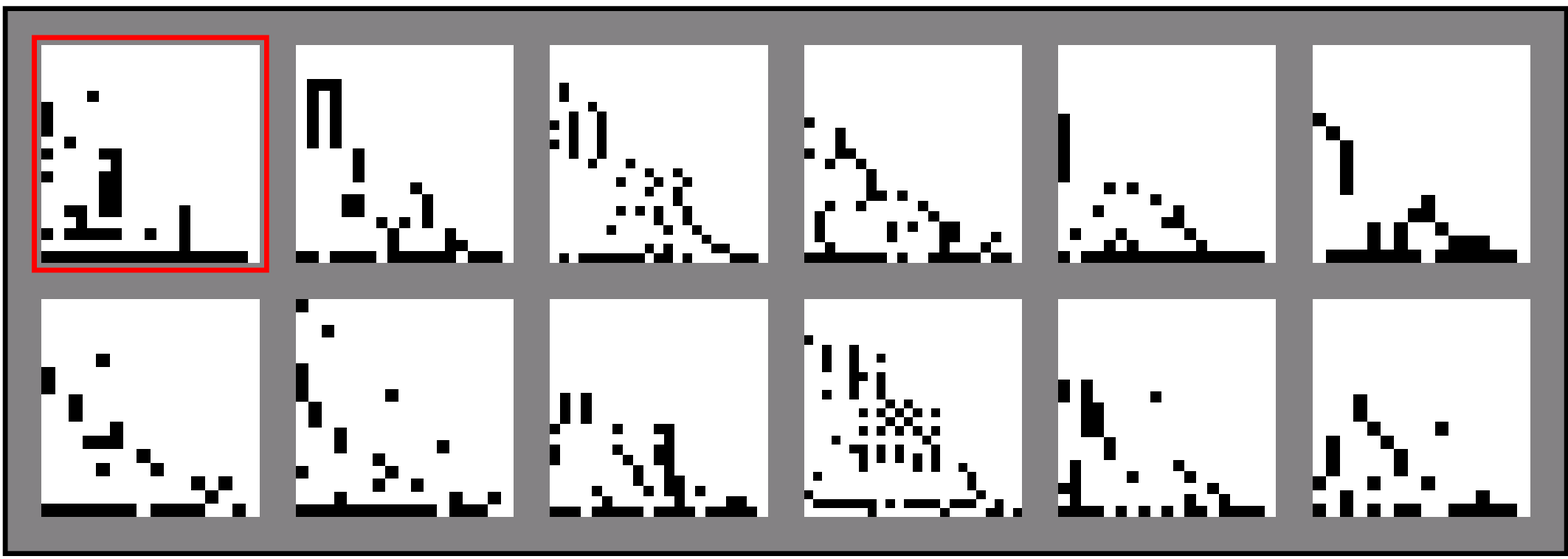}\\

\smallskip

  \includegraphics[width=2\figurewidth,keepaspectratio,clip]{Canton-RY.eps}
  \label{fig:canton}
\end{figure}
\begin{figure}[h]
  \caption{Caribbean Reef (see \ref{sec:more} for explanations)}
  \textsf{Matching model}:\\
  \includegraphics[width=2\figurewidth,keepaspectratio,clip]{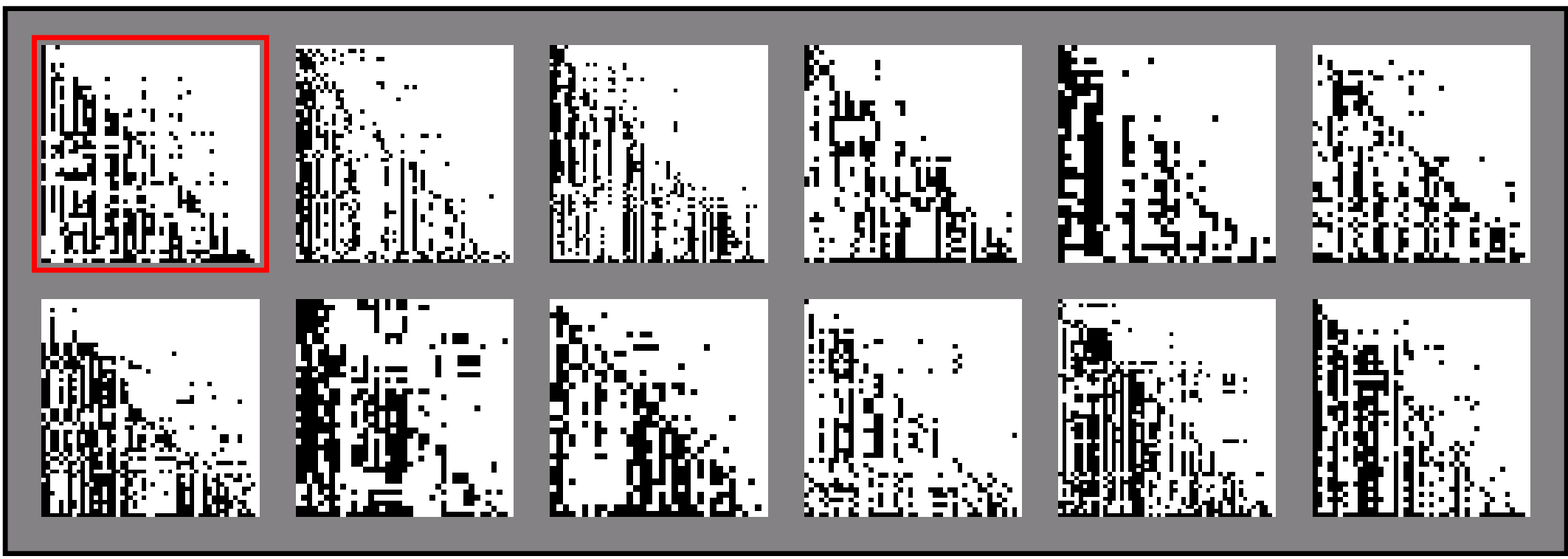}\\
  \textsf{Niche model}:\\
  \includegraphics[width=2\figurewidth,keepaspectratio,clip]{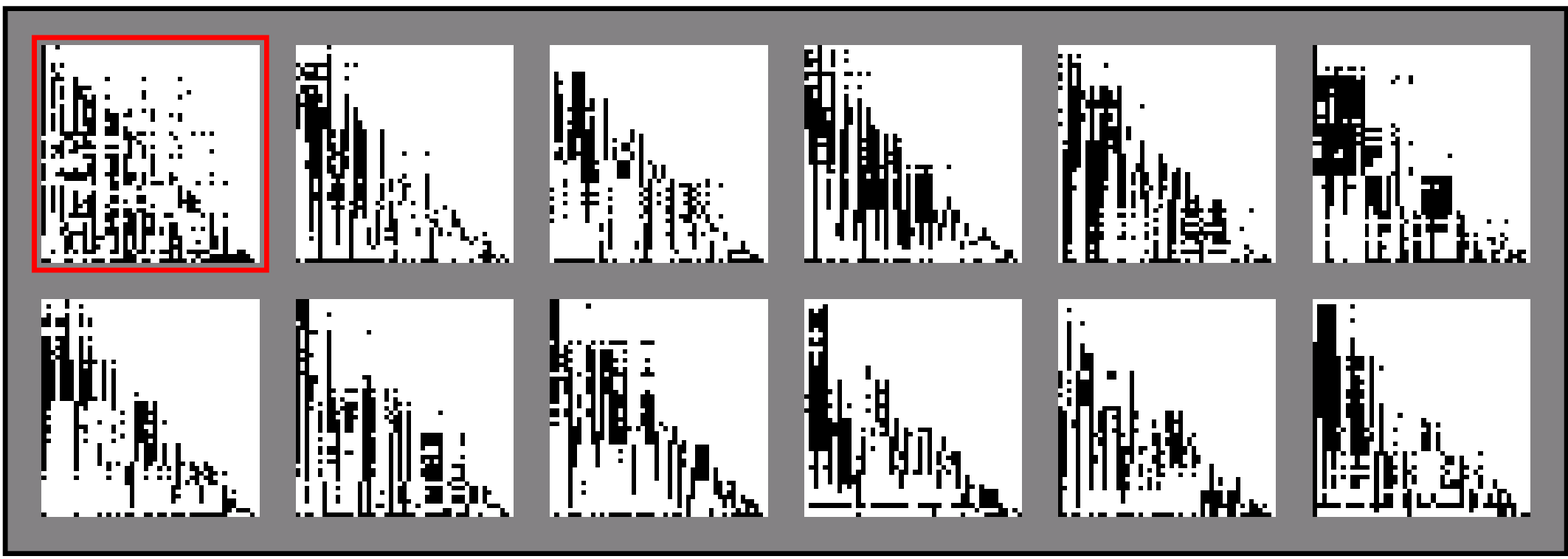}\\

\smallskip

  \includegraphics[width=2\figurewidth,keepaspectratio,clip]{Reef-small-RY.eps}
  \label{fig:reef}
\end{figure}
\begin{figure}[h]
  \caption{Chesapeake Bay (see \ref{sec:more} for explanations)}
  \textsf{Matching model}:\\
  \includegraphics[width=2\figurewidth,keepaspectratio,clip]{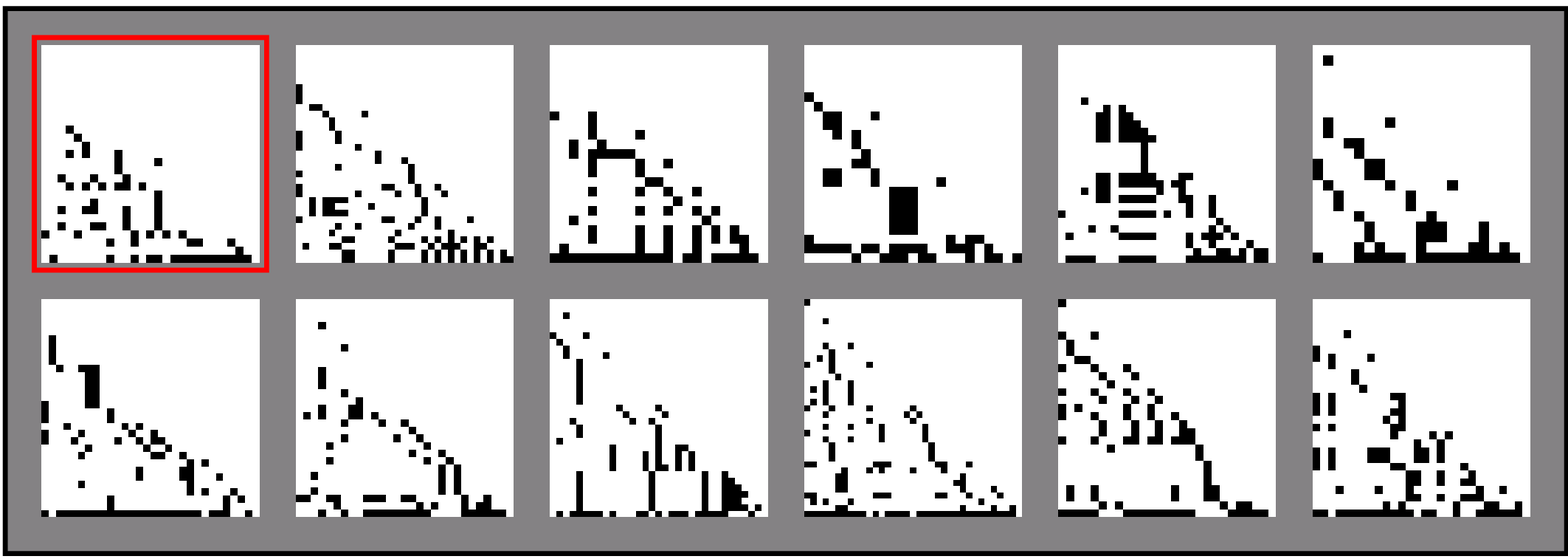}\\
  \textsf{Niche model}:\\
  \includegraphics[width=2\figurewidth,keepaspectratio,clip]{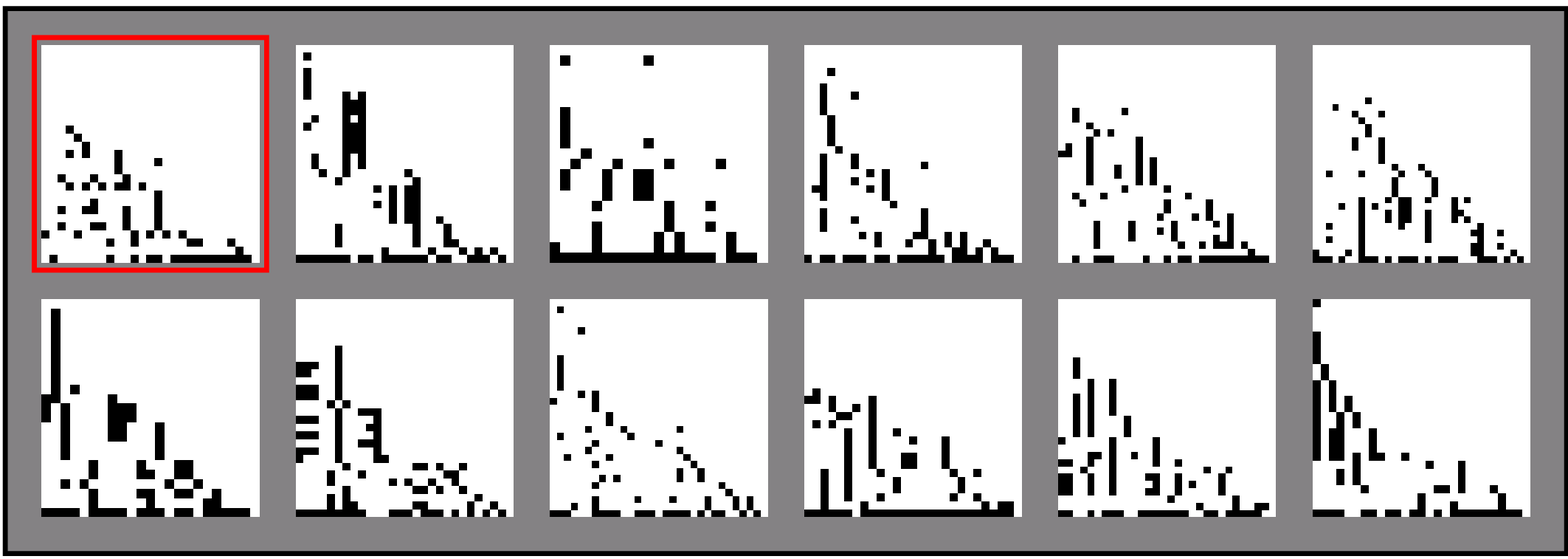}\\

\smallskip

  \includegraphics[width=2\figurewidth,keepaspectratio,clip]{Chesapeake-RY.eps}
  \label{fig:cheaspeake}
\end{figure}
\begin{figure}[h]
  \caption{Coachella Valley (see \ref{sec:more} for explanations)}
  \textsf{Matching model}:\\
  \includegraphics[width=2\figurewidth,keepaspectratio,clip]{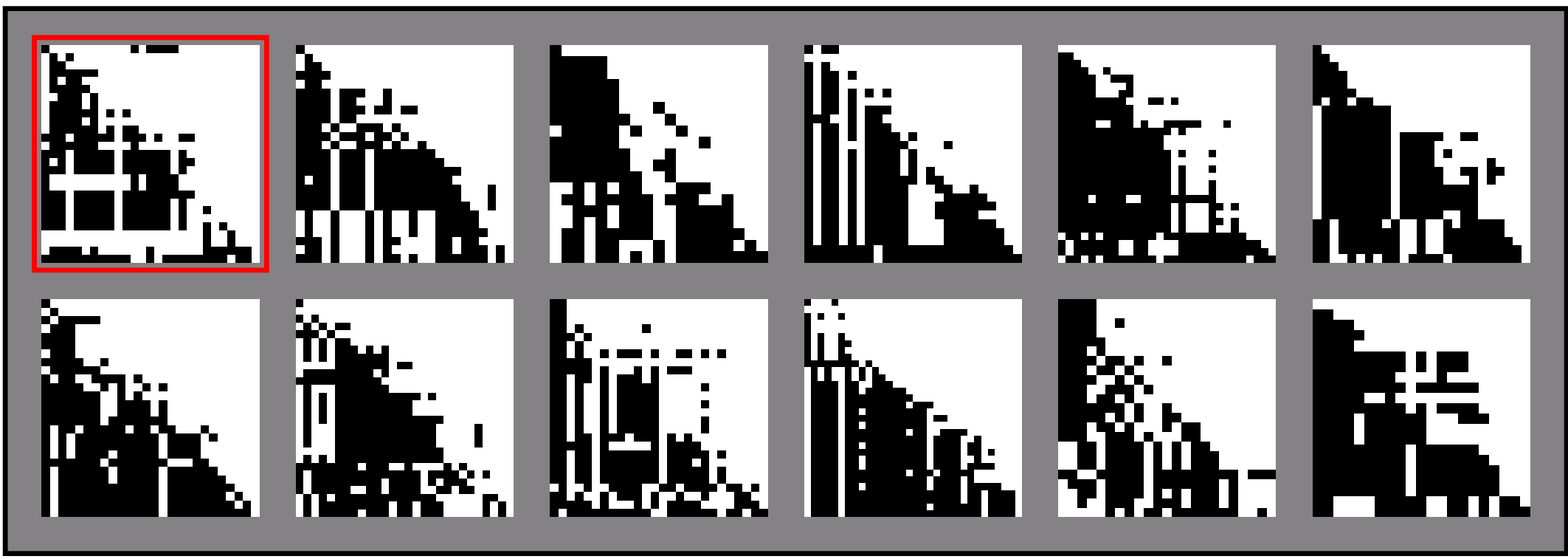}\\
  \textsf{Niche model}:\\
  \includegraphics[width=2\figurewidth,keepaspectratio,clip]{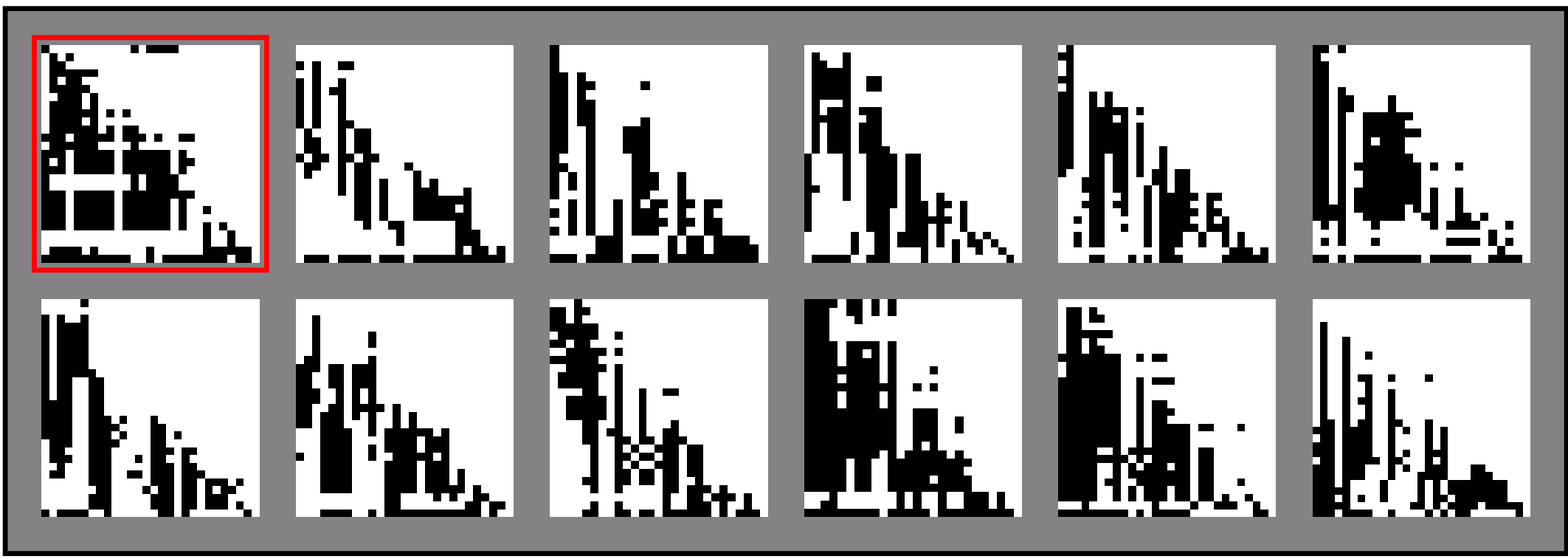}\\

\smallskip

  \includegraphics[width=2\figurewidth,keepaspectratio,clip]{Coachella-RY.eps}
  \label{fig:coachella}
\end{figure}
\begin{figure}[h]
  \caption{El Verde Rainforest (see \ref{sec:more} for explanations)}
  \textsf{Matching model}:\\
  \includegraphics[width=2\figurewidth,keepaspectratio,clip]{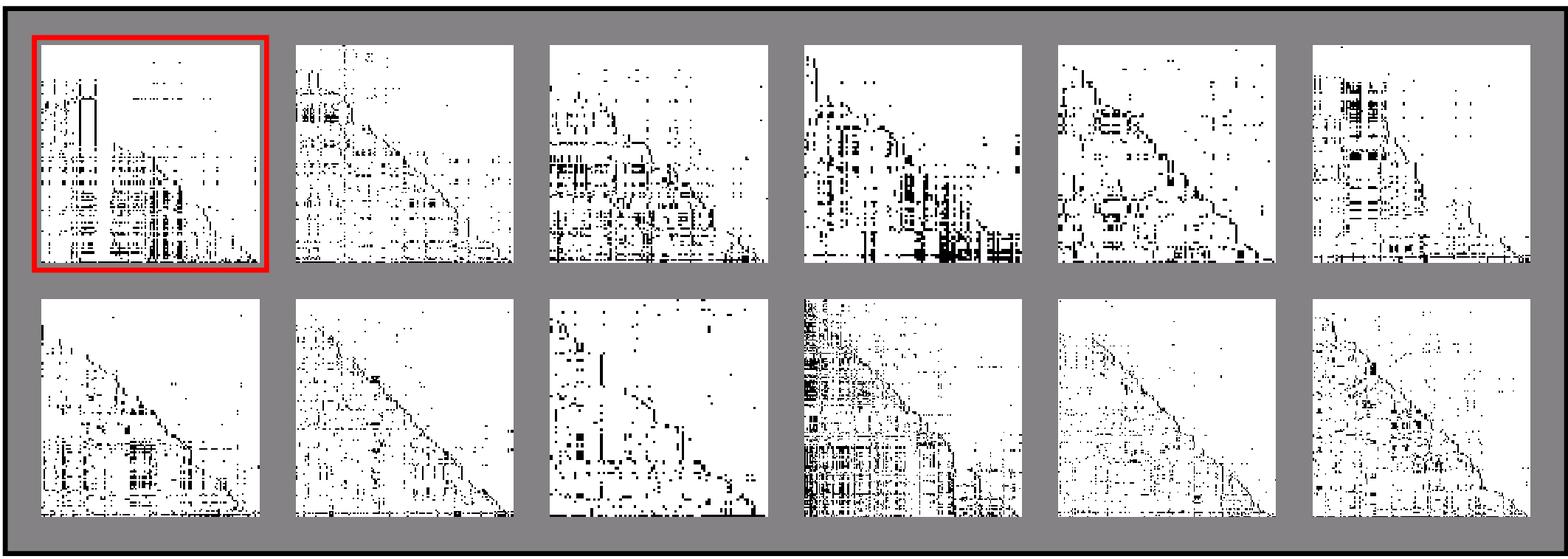}\\
  \textsf{Niche model}:\\
  \includegraphics[width=2\figurewidth,keepaspectratio,clip]{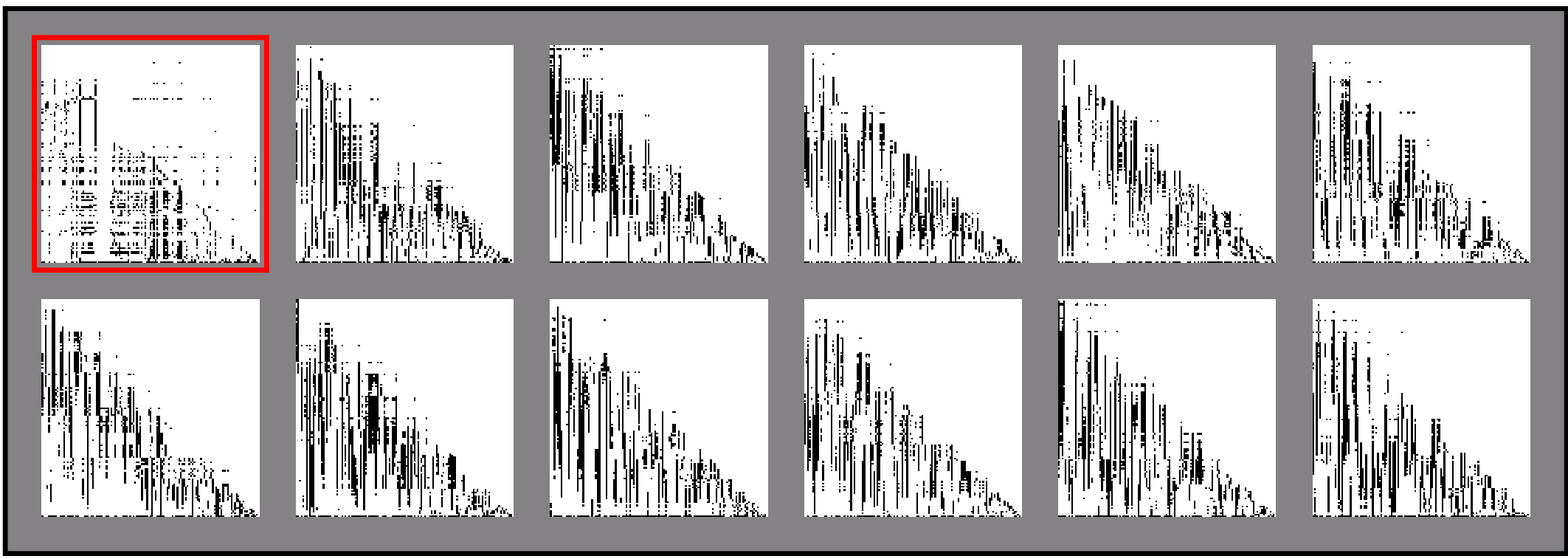}\\

\smallskip

  \includegraphics[width=2\figurewidth,keepaspectratio,clip]{El-Verde-RY.eps}
  \label{fig:verde}
\end{figure}
\begin{figure}[h]
  \caption{Little Rock Lake (see \ref{sec:more} for explanations)}
  \textsf{Matching model}:\\
  \includegraphics[width=2\figurewidth,keepaspectratio,clip]{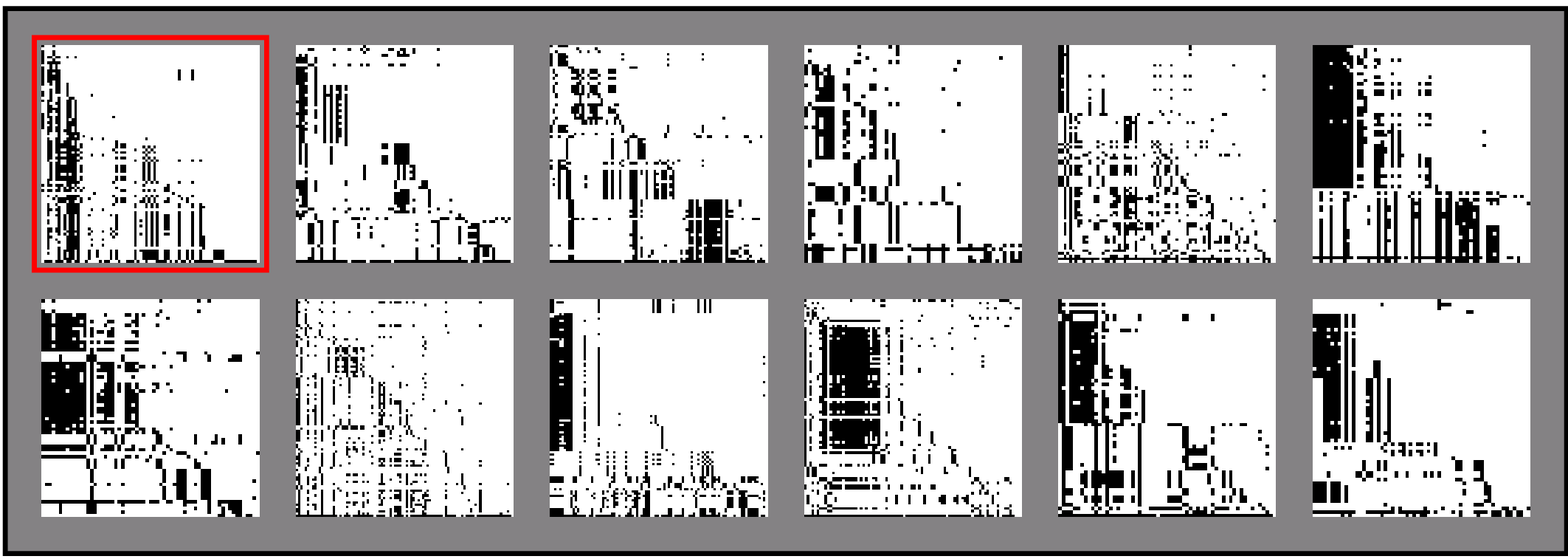}\\
  \textsf{Niche model}:\\
  \includegraphics[width=2\figurewidth,keepaspectratio,clip]{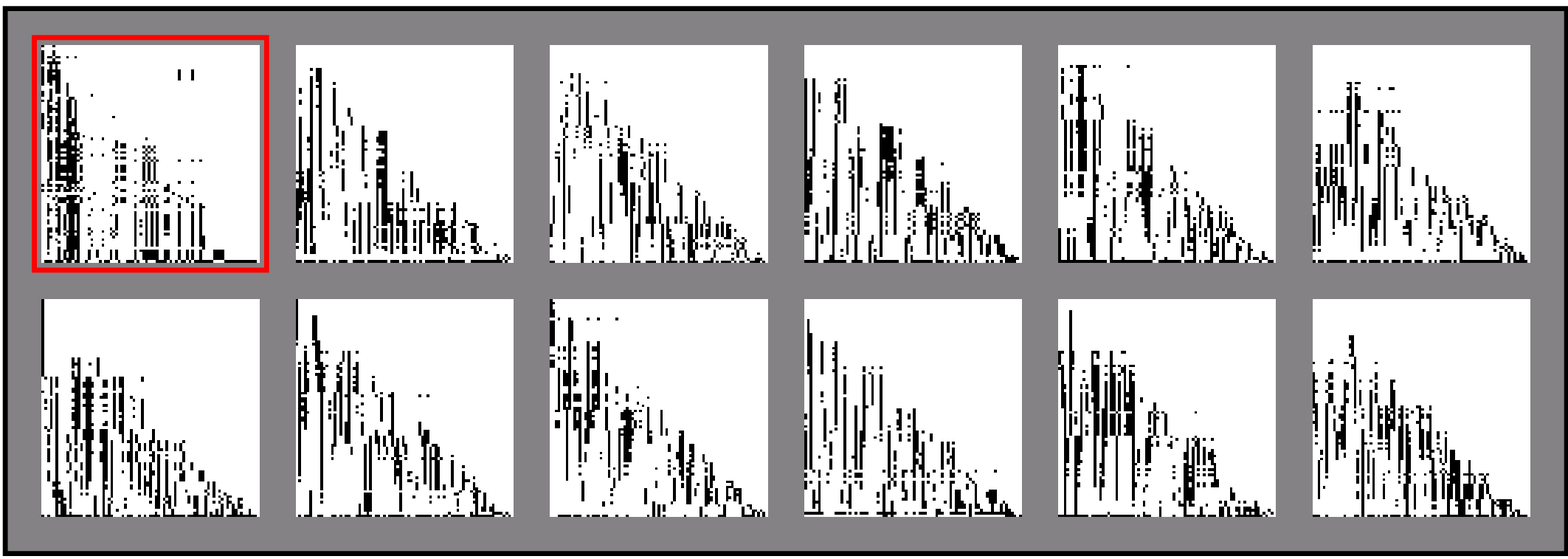}\\

\smallskip

  \includegraphics[width=2\figurewidth,keepaspectratio,clip]{Little-Rock-RY.eps}
  \label{fig:rock}
\end{figure}
\begin{figure}[h]
  \caption{Northeast US Shelf (see \ref{sec:more} for explanations)}
  \textsf{Matching model}:\\
  \includegraphics[width=2\figurewidth,keepaspectratio,clip]{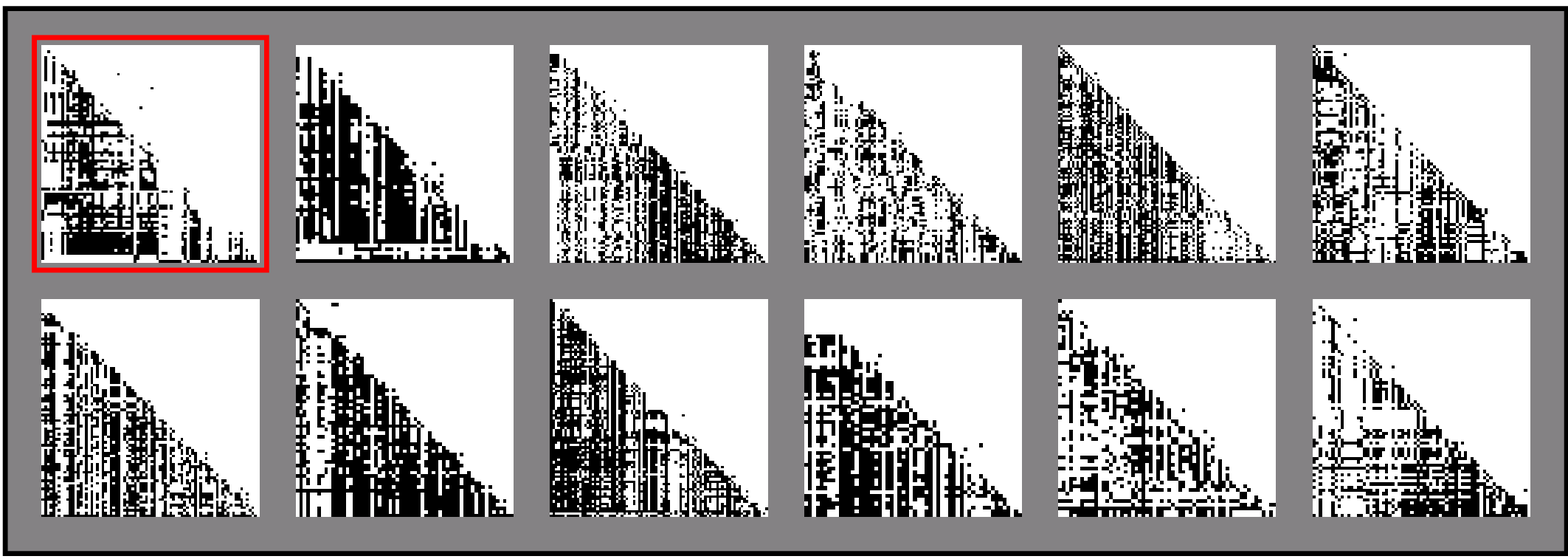}\\
  \textsf{Niche model}:\\
  \includegraphics[width=2\figurewidth,keepaspectratio,clip]{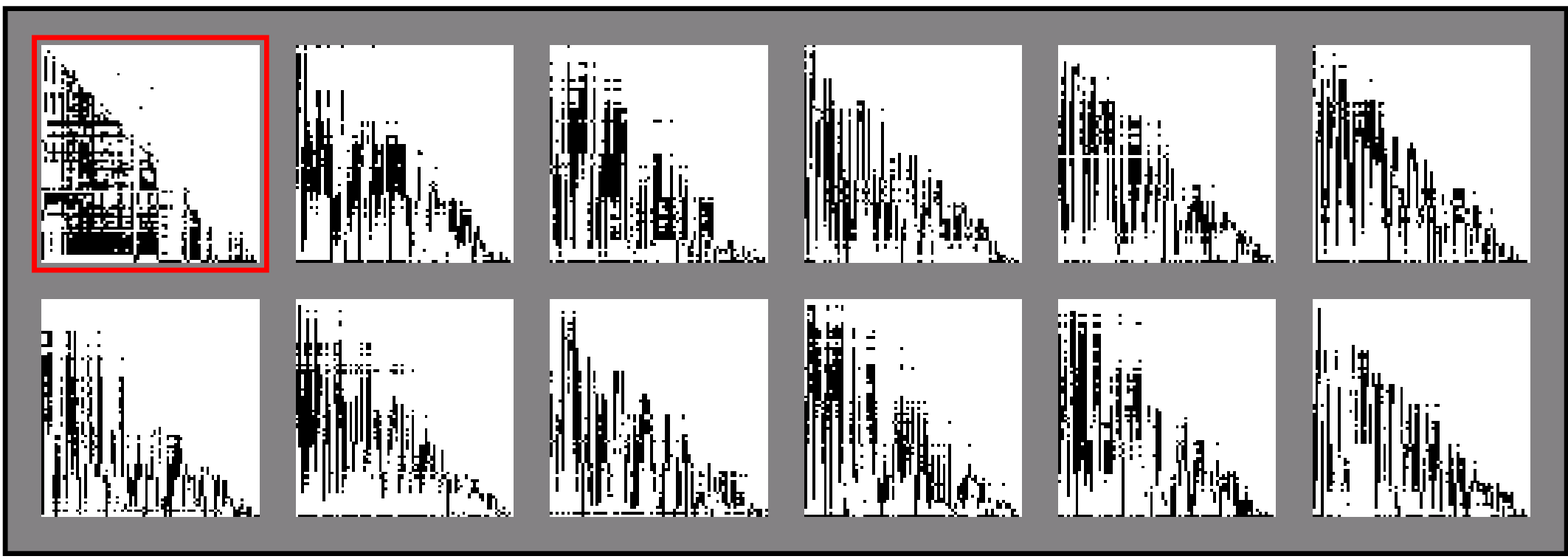}\\

\smallskip

  \includegraphics[width=2\figurewidth,keepaspectratio,clip]{Shelf-RY.eps}
  \label{fig:shelf}
\end{figure}
\begin{figure}[h]
  \caption{Scotch Broom (see \ref{sec:more} for explanations)}
  \textsf{Matching model}:\\
  \includegraphics[width=2\figurewidth,keepaspectratio,clip]{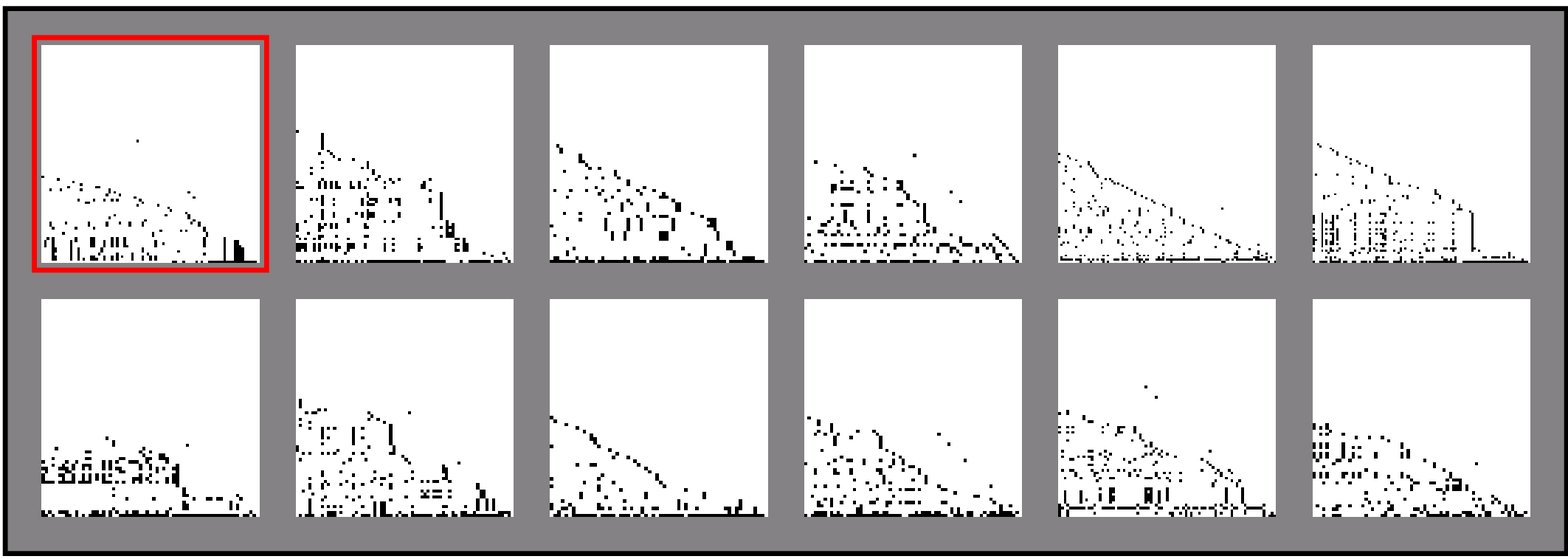}\\
  \textsf{Niche model}:\\
  \includegraphics[width=2\figurewidth,keepaspectratio,clip]{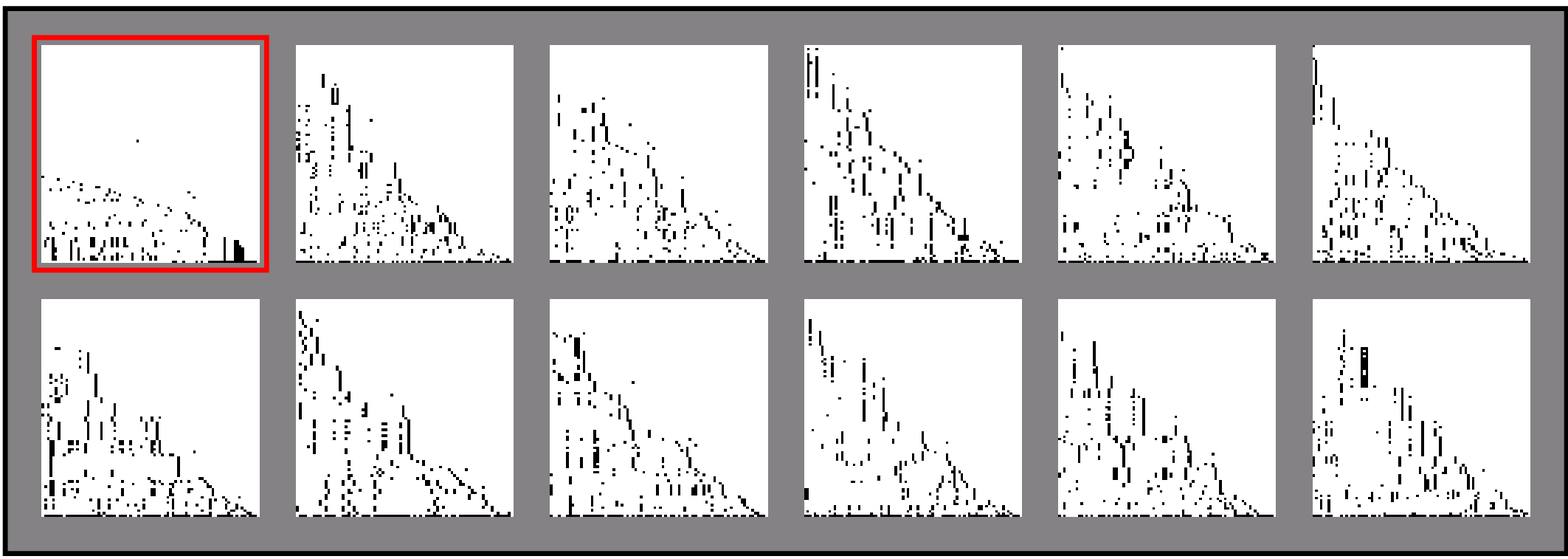}\\

\smallskip

  \includegraphics[width=2\figurewidth,keepaspectratio,clip]{Broom-RY.eps}
  \label{fig:broom}
\end{figure}
\begin{figure}[h]
  \caption{Skipwith Pond (see \ref{sec:more} for explanations)}
  \textsf{Matching model}:\\
  \includegraphics[width=2\figurewidth,keepaspectratio,clip]{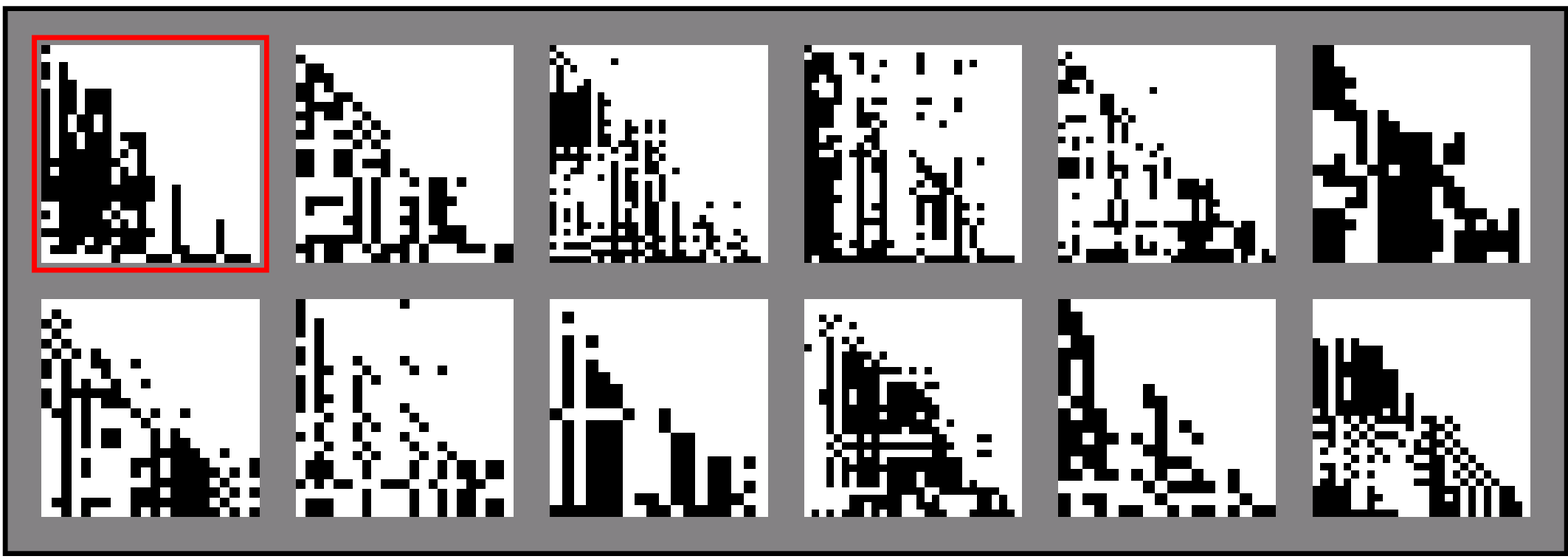}\\
  \textsf{Niche model}:\\
  \includegraphics[width=2\figurewidth,keepaspectratio,clip]{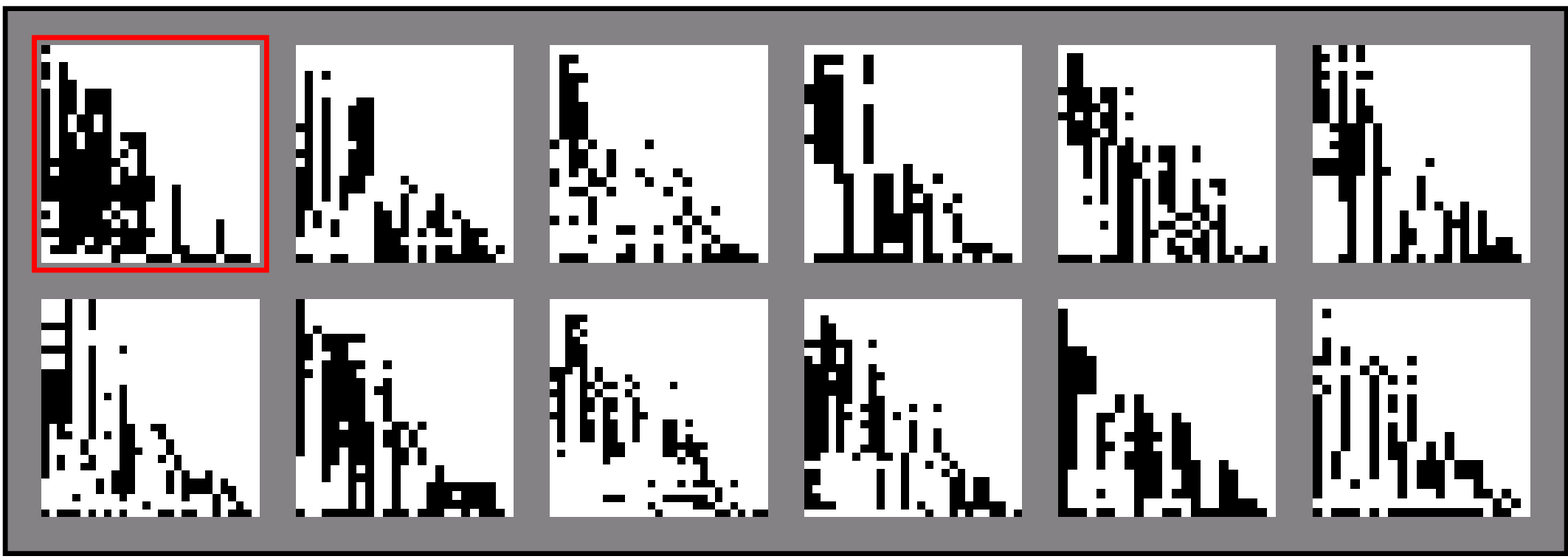}\\

\smallskip

  \includegraphics[width=2\figurewidth,keepaspectratio,clip]{Skipwith-RY.eps}
  \label{fig:skipwith}
\end{figure}
\begin{figure}[h]
  \caption{St. Marks Seegrass (see \ref{sec:more} for explanations)}
  \textsf{Matching model}:\\
  \includegraphics[width=2\figurewidth,keepaspectratio,clip]{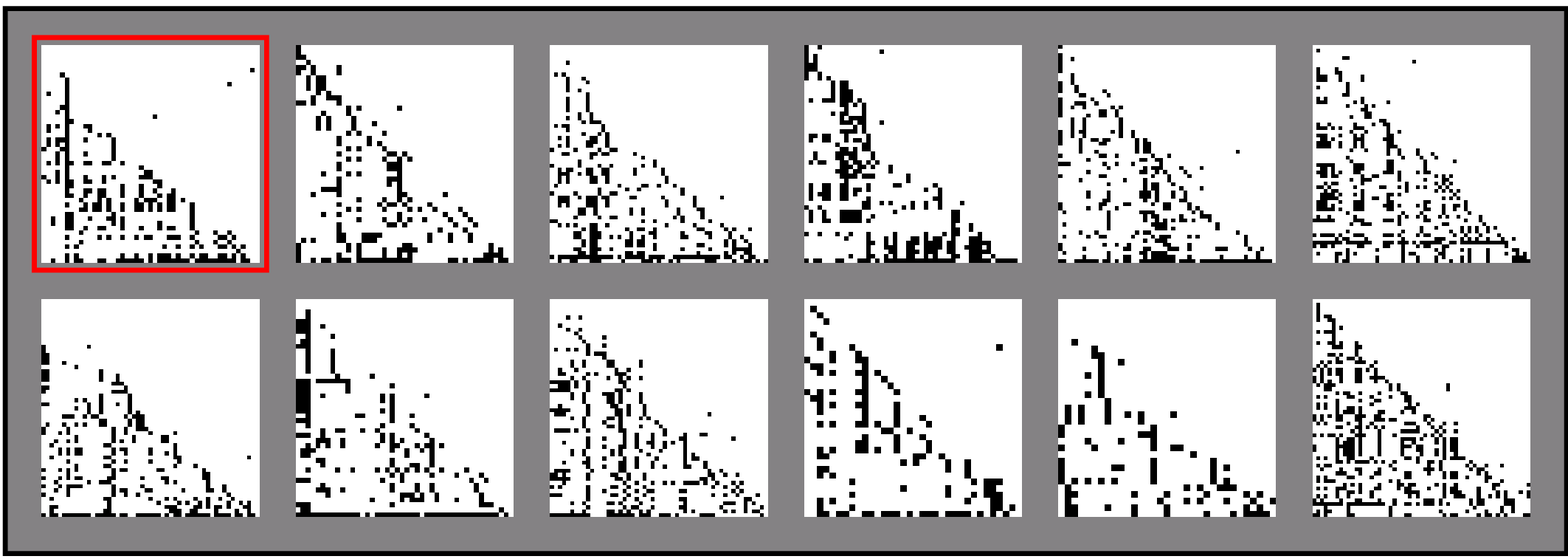}\\
  \textsf{Niche model}:\\
  \includegraphics[width=2\figurewidth,keepaspectratio,clip]{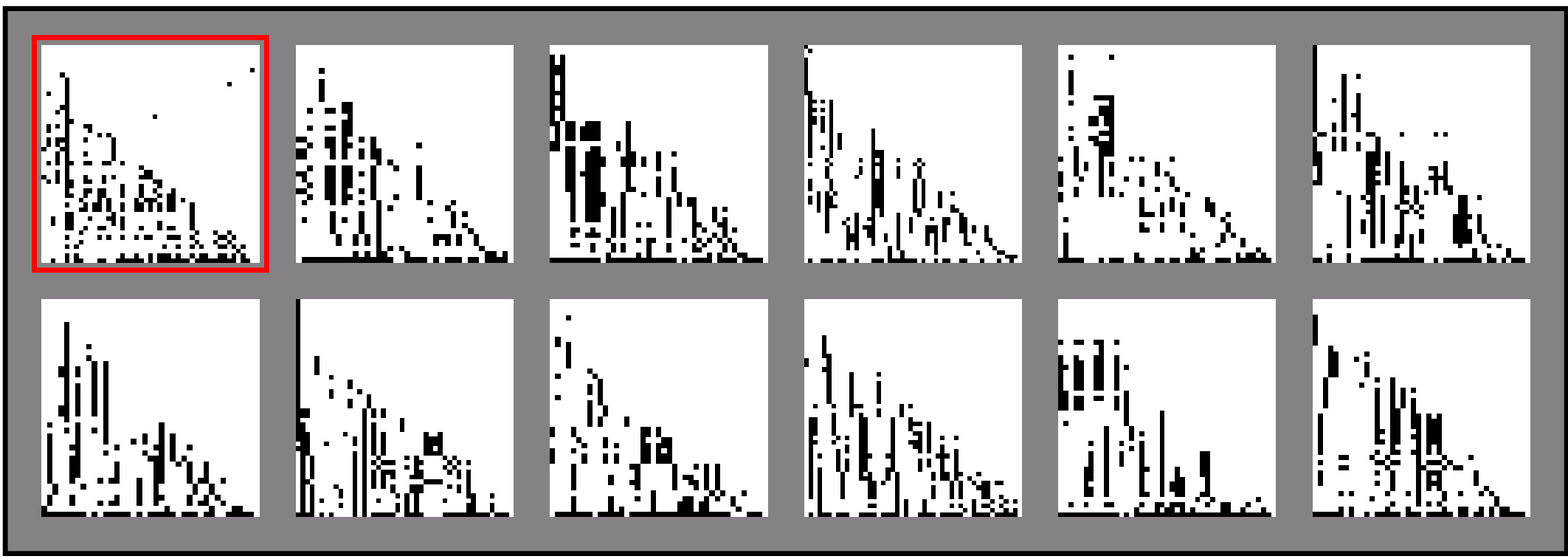}\\

\smallskip

  \includegraphics[width=2\figurewidth,keepaspectratio,clip]{St-Marks-RY.eps}
  \label{fig:marks}
\end{figure}
\begin{figure}[h]
  \caption{St. Martin Island (see \ref{sec:more} for explanations)}
  \textsf{Matching model}:\\
  \includegraphics[width=2\figurewidth,keepaspectratio,clip]{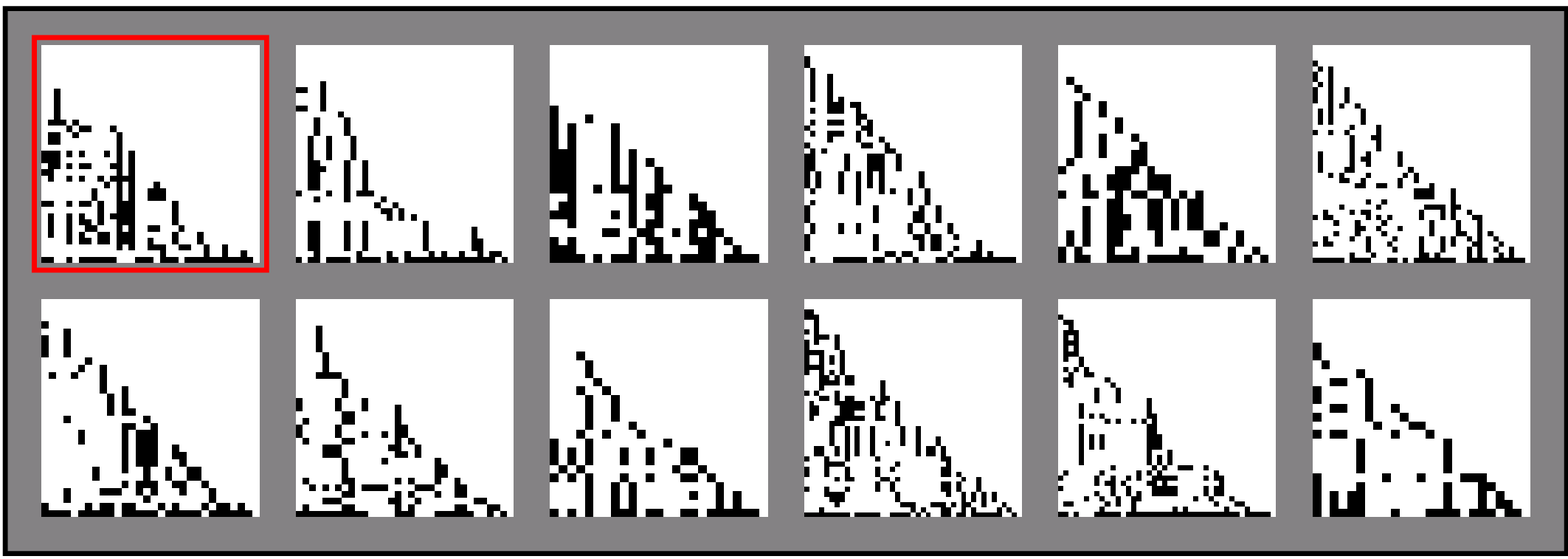}\\
  \textsf{Niche model}:\\
  \includegraphics[width=2\figurewidth,keepaspectratio,clip]{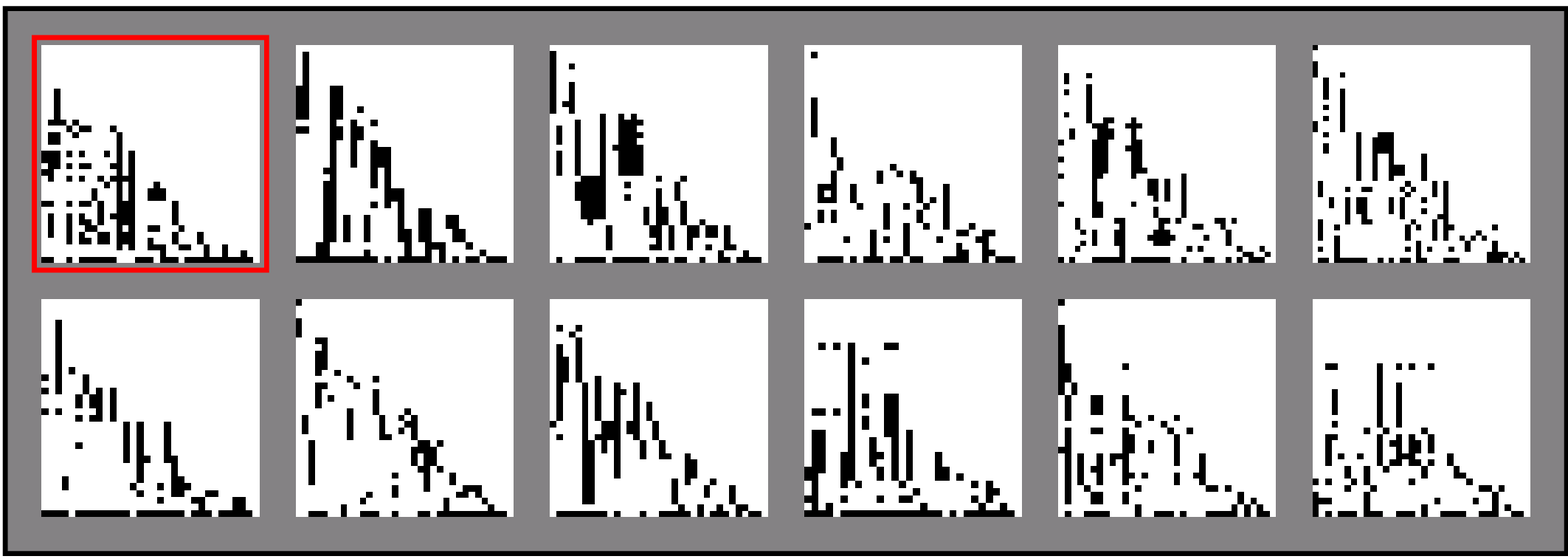}\\

\smallskip

  \includegraphics[width=2\figurewidth,keepaspectratio,clip]{St-Martin-RY.eps}
  \label{fig:martin}
\end{figure}
\begin{figure}[h]
  \caption{Stony Stream (see \ref{sec:more} for explanations)}
  \textsf{Matching model}:\\
  \includegraphics[width=2\figurewidth,keepaspectratio,clip]{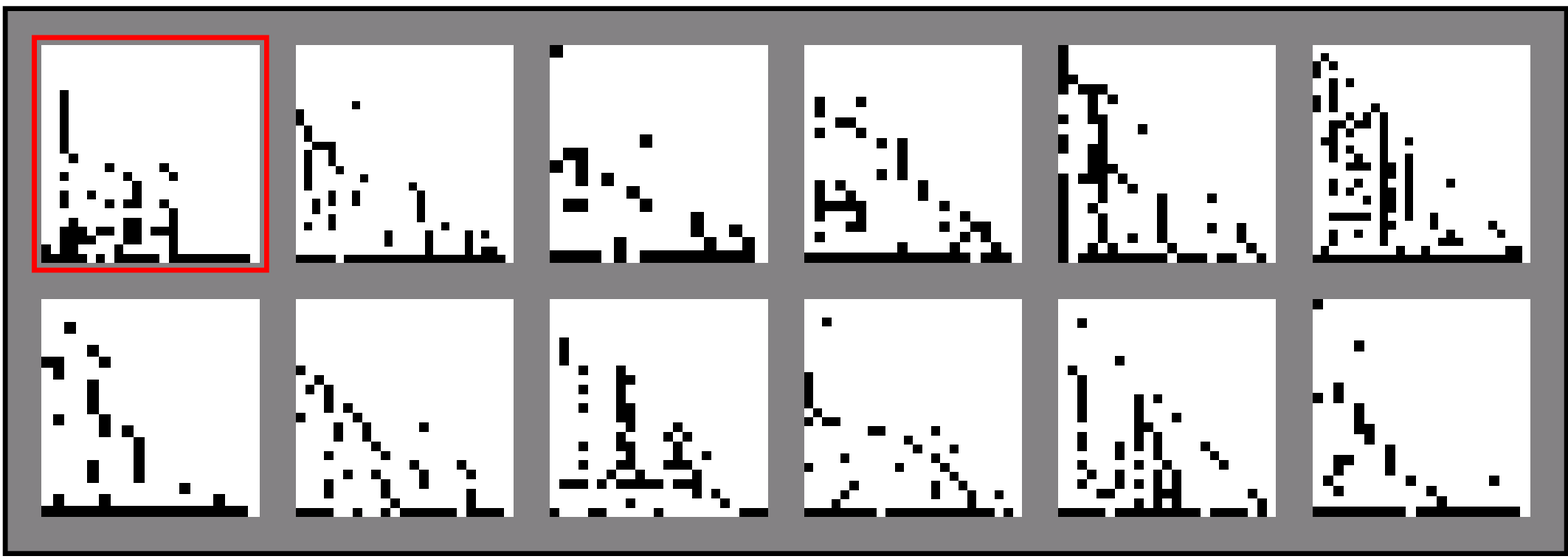}\\
  \textsf{Niche model}:\\
  \includegraphics[width=2\figurewidth,keepaspectratio,clip]{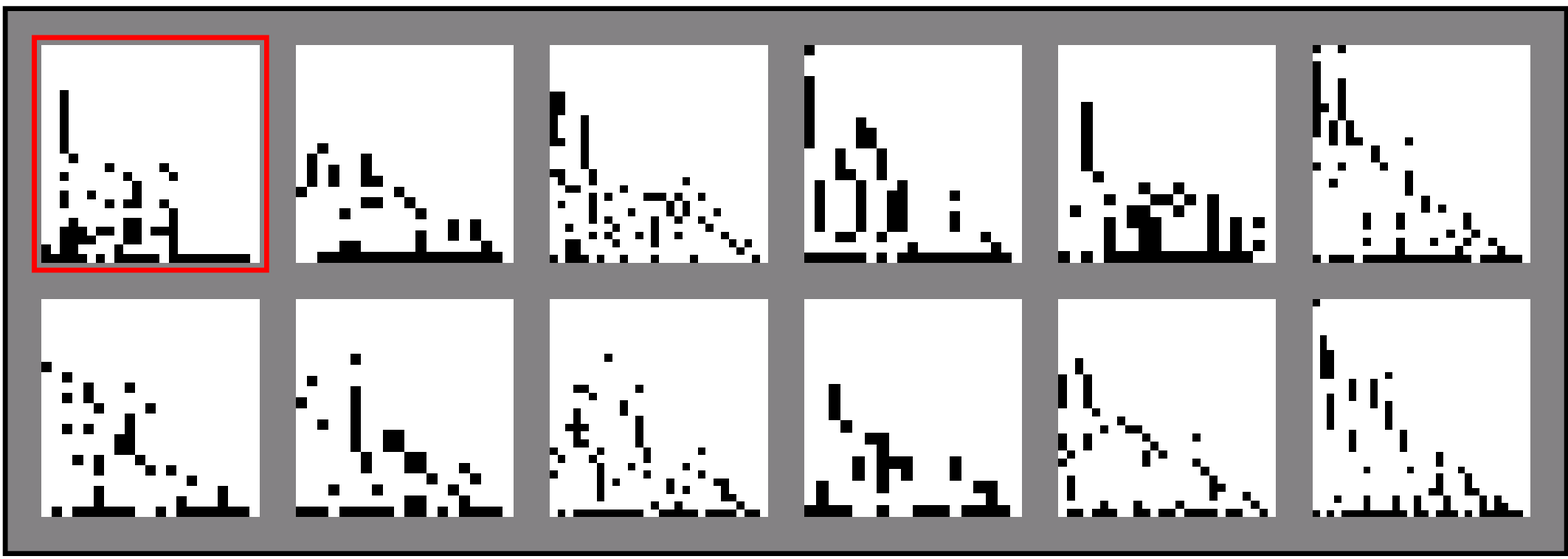}\\

\smallskip

  \includegraphics[width=2\figurewidth,keepaspectratio,clip]{Stony-RY.eps}
  \label{fig:stony}
\end{figure}
\begin{figure}[h]
  \caption{Ythan Estuary 1 (see \ref{sec:more} for explanations)}
  \textsf{Matching model}:\\
  \includegraphics[width=2\figurewidth,keepaspectratio,clip]{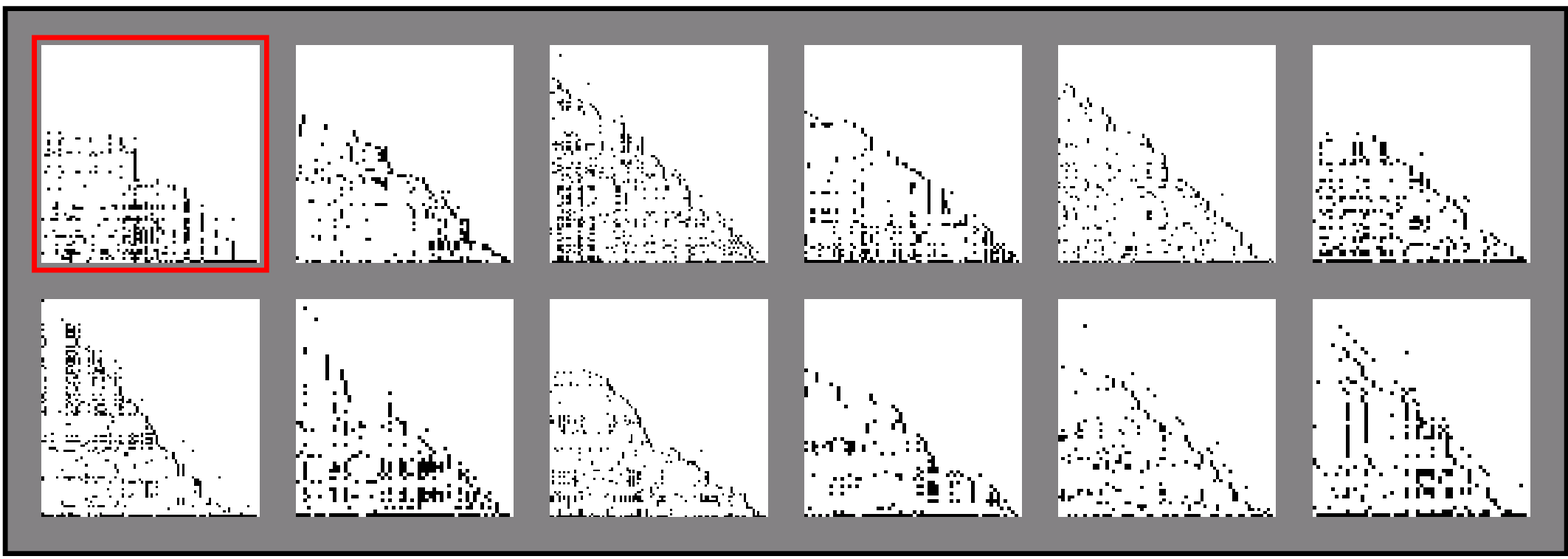}\\
  \textsf{Niche model}:\\
  \includegraphics[width=2\figurewidth,keepaspectratio,clip]{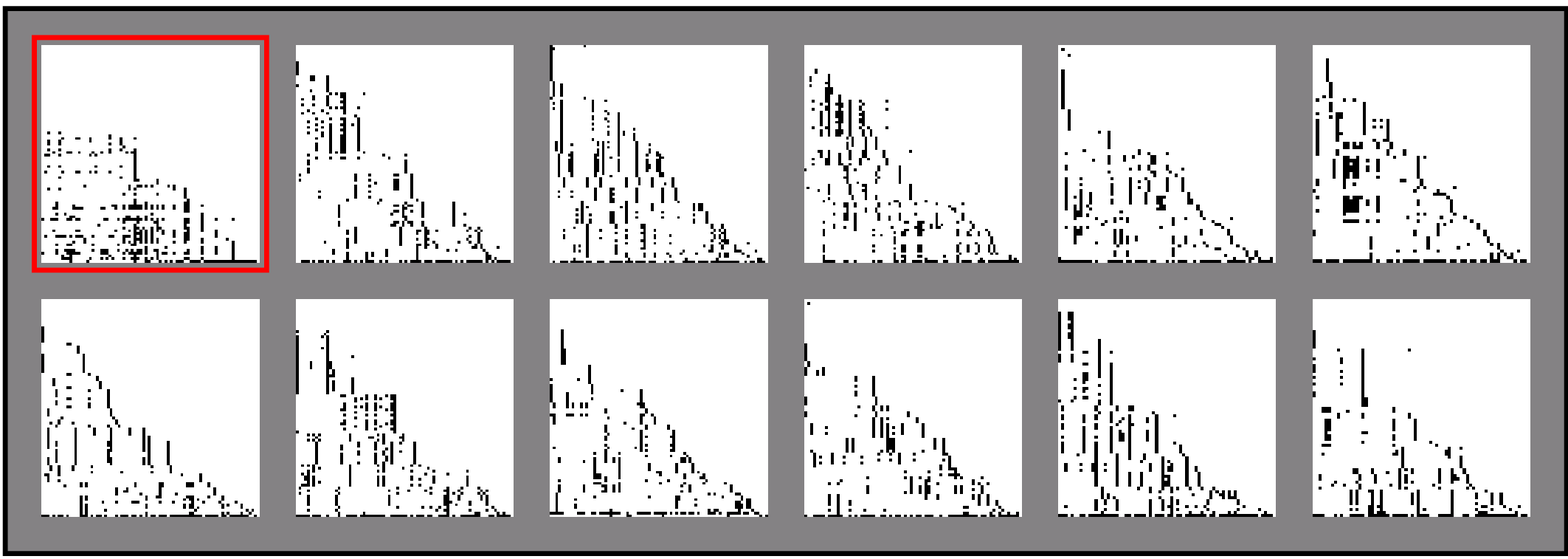}\\

\smallskip

  \includegraphics[width=2\figurewidth,keepaspectratio,clip]{Ythan91-RY.eps}
  \label{fig:ythan91}
\end{figure}
\begin{figure}[h]
  \caption{Ythan Estuary 2 (see \ref{sec:more} for explanations)}
  \textsf{Matching model}:\\
  \includegraphics[width=2\figurewidth,keepaspectratio,clip]{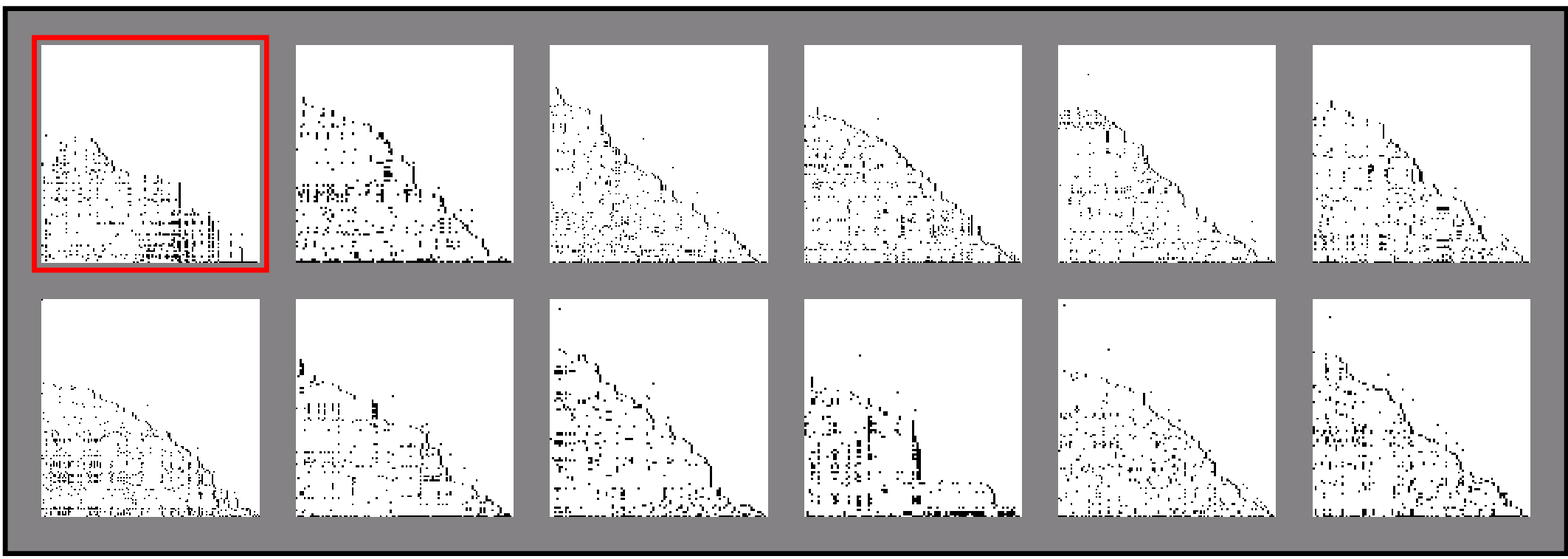}\\
  \textsf{Niche model}:\\
  \includegraphics[width=2\figurewidth,keepaspectratio,clip]{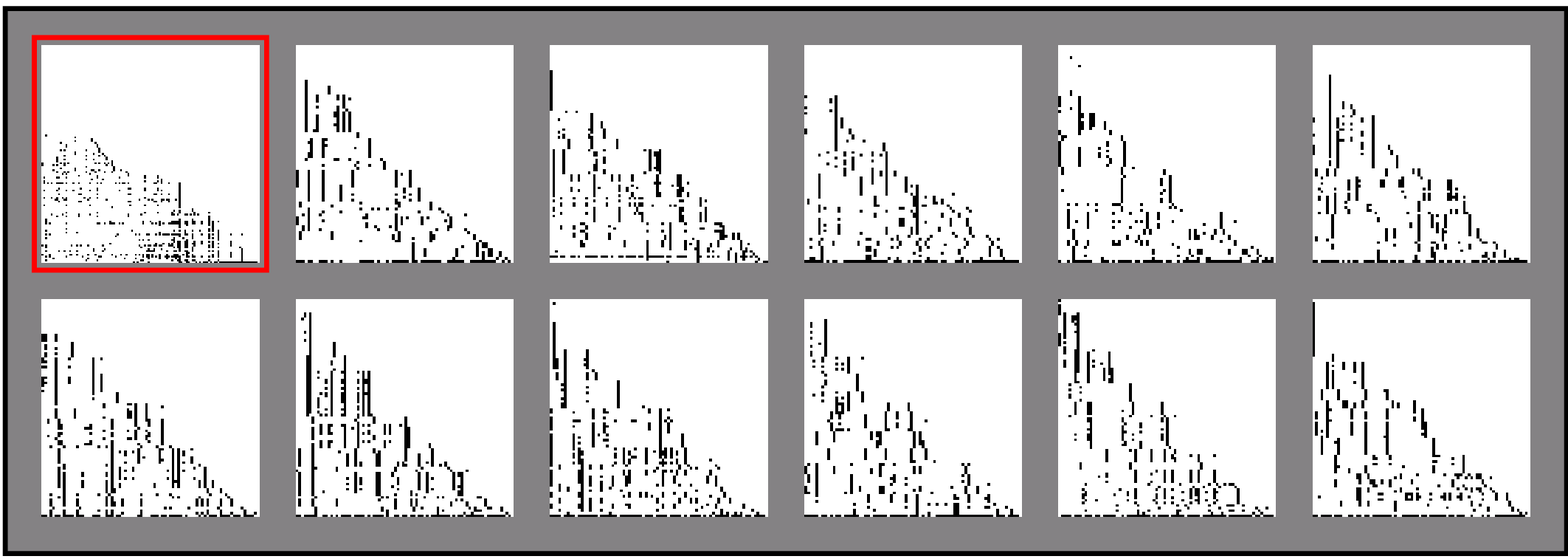}\\

\smallskip

  \includegraphics[width=2\figurewidth,keepaspectratio,clip]{Ythan96-RY.eps}
  \label{fig:ythan96}
\end{figure}

\clearpage


\end{bibunit}

\excludeversion{never}
\begin{never}
  \bibliography{/home/axel/bib/bibview}
\end{never}

\end{nowordcount}
\end{document}

